\documentclass[12pt,preprint]{aastex}

\shorttitle{SNe in CS Bubbles}
\shortauthors{Dwarkadas.}
\begin{document} 
\newcommand{\vper}{\mbox{${v_{\perp}}$}}
\newcommand{\vpar}{\mbox{${v_{\parallel}}$}}
\newcommand{\uper}{\mbox{${u_{\perp}}$}}
\newcommand{\vperout}{\mbox{${{v_{\perp}}_{o}}$}}
\newcommand{\uperout}{\mbox{${{u_{\perp}}_{o}}$}}
\newcommand{\vperin}{\mbox{${{v_{\perp}}_{i}}$}}
\newcommand{\uperin}{\mbox{${{u_{\perp}}_{i}}$}}
\newcommand{\upar}{\mbox{${u_{\parallel}}$}}
\newcommand{\uparout}{\mbox{${{u_{\parallel}}_{o}}$}}
\newcommand{\vparout}{\mbox{${{v_{\parallel}}_{o}}$}}
\newcommand{\uparin}{\mbox{${{u_{\parallel}}_{i}}$}}
\newcommand{\vparin}{\mbox{${{v_{\parallel}}_{i}}$}}
\newcommand{\dout}{\mbox{${\rho}_{o}$}}
\newcommand{\din}{\mbox{${\rho}_{i}$}}
\newcommand{\da}{\mbox{${\rho}_{1}$}}
\newcommand{\mfast}{\mbox{$\dot{M}_{f}$}}
\newcommand{\mslow}{\mbox{$\dot{M}_{a}$}}
\newcommand{\beqn}{\begin{eqnarray}}
\newcommand{\eeqn}{\end{eqnarray}}
\newcommand{\be}{\begin{equation}}
\newcommand{\ee}{\end{equation}}
\newcommand{\noi}{\noindent}
\newcommand{\ftheta}{\mbox{$f(\theta)$}}
\newcommand{\gtheta}{\mbox{$g(\theta)$}}
\newcommand{\ltheta}{\mbox{$L(\theta)$}}
\newcommand{\stheta}{\mbox{$S(\theta)$}}
\newcommand{\utheta}{\mbox{$U(\theta)$}}
\newcommand{\xitheta}{\mbox{$\xi(\theta)$}}
\newcommand{\vs}{\mbox{${v_{s}}$}}
\newcommand{\ro}{\mbox{${R_{0}}$}}
\newcommand{\pa}{\mbox{${P_{1}}$}}
\newcommand{\va}{\mbox{${v_{a}}$}}
\newcommand{\vo}{\mbox{${v_{o}}$}}
\newcommand{\vp}{\mbox{${v_{p}}$}}
\newcommand{\vw}{\mbox{${v_{w}}$}}
\newcommand{\vf}{\mbox{${v_{f}}$}}
\newcommand{\lprime}{\mbox{${L^{\prime}}$}}
\newcommand{\uprime}{\mbox{${U^{\prime}}$}}
\newcommand{\sprime}{\mbox{${S^{\prime}}$}}
\newcommand{\xiprime}{\mbox{${{\xi}^{\prime}}$}}
\newcommand{\mdot}{\mbox{$\dot{M}$}}
\newcommand{\msun}{\mbox{$M_{\odot}$}}
\newcommand{\yr}{\mbox{${\rm yr}^{-1}$}}
\newcommand{\kms}{\mbox{${\rm km} \;{\rm s}^{-1}$}}
\newcommand{\lambdav}{\mbox{${\lambda}_{v}$}}
\newcommand{\lequ}{\mbox{${L_{eq}}$}}
\newcommand{\eqpratio}{\mbox{${R_{eq}/R_{p}}$}}
\newcommand{\ra}{\mbox{${r_{o}}$}}
\newcommand{\bfig}{\begin{figure}[h]}
\newcommand{\efig}{\end{figure}}
\newcommand{\tone}{\mbox{${t_{1}}$}}
\newcommand{\done}{\mbox{${{\rho}_{1}}$}}
\newcommand{\dsn}{\mbox{${\rho}_{SN}$}}
\newcommand{\dzero}{\mbox{${\rho}_{0}$}}
\newcommand{\ve}{\mbox{${v}_{e}$}}
\newcommand{\vej}{\mbox{${v}_{ej}$}}
\newcommand{\Mch}{\mbox{${M}_{ch}$}}
\newcommand{\mej}{\mbox{${M}_{e}$}}
\newcommand{\Mst}{\mbox{${M}_{ST}$}}
\newcommand{\dam}{\mbox{${\rho}_{am}$}}
\newcommand{\Rst}{\mbox{${R}_{ST}$}}
\newcommand{\Vst}{\mbox{${V}_{ST}$}}
\newcommand{\Tst}{\mbox{${T}_{ST}$}}
\newcommand{\no}{\mbox{${n}_{0}$}}
\newcommand{\Efif}{\mbox{${E}_{51}$}}
\newcommand{\rsh}{\mbox{${R}_{sh}$}}
\newcommand{\msh}{\mbox{${M}_{sh}$}}
\newcommand{\vsh}{\mbox{${V}_{sh}$}}
\newcommand{\vrev}{\mbox{${v}_{rev}$}}
\newcommand{\rpr}{\mbox{${R}^{\prime}$}}
\newcommand{\mpr}{\mbox{${M}^{\prime}$}}
\newcommand{\vpr}{\mbox{${V}^{\prime}$}}
\newcommand{\tpr}{\mbox{${t}^{\prime}$}}
\newcommand{\cone}{\mbox{${c}_{1}$}}
\newcommand{\ctwo}{\mbox{${c}_{2}$}}
\newcommand{\cthree}{\mbox{${c}_{3}$}}
\newcommand{\cfour}{\mbox{${c}_{4}$}}
\newcommand{\Te}{\mbox{${T}_{e}$}}
\newcommand{\Ti}{\mbox{${T}_{i}$}}
\newcommand{\Ha}{\mbox{${H}_{\alpha}$}}
\newcommand{\Rprime}{\mbox{${R}^{\prime}$}}
\newcommand{\Vprime}{\mbox{${V}^{\prime}$}}
\newcommand{\Tprime}{\mbox{${T}^{\prime}$}}
\newcommand{\Mprime}{\mbox{${M}^{\prime}$}}
\newcommand{\rprime}{\mbox{${r}^{\prime}$}}
\newcommand{\rfprime}{\mbox{${r}_f^{\prime}$}}
\newcommand{\vprime}{\mbox{${v}^{\prime}$}}
\newcommand{\tprime}{\mbox{${t}^{\prime}$}}
\newcommand{\mprime}{\mbox{${m}^{\prime}$}}
\newcommand{\Me}{\mbox{${M}_{e}$}}
\newcommand{\nh}{\mbox{${n}_{H}$}}
\newcommand{\rr}{\mbox{${R}_{2}$}}
\newcommand{\rf}{\mbox{${R}_{1}$}}
\newcommand{\vtwo}{\mbox{${V}_{2}$}}
\newcommand{\vout}{\mbox{${V}_{1}$}}
\newcommand{\dshell}{\mbox{${{\rho}_{sh}}$}}
\newcommand{\dwind}{\mbox{${{\rho}_{w}}$}}
\newcommand{\dslow}{\mbox{${{\rho}_{s}}$}}
\newcommand{\dfast}{\mbox{${{\rho}_{f}}$}}
\newcommand{\vfast}{\mbox{${v}_{f}$}}
\newcommand{\vslow}{\mbox{${v}_{s}$}}
\newcommand{\cc}{\mbox{${\rm cm}^{-3}$}}
 
\title{The Evolution of Supernovae in Circumstellar Wind Bubbles II:
  Case of a Wolf-Rayet star}

\author{ Vikram V. Dwarkadas}

\affil{Astronomy and Astrophysics, Univ of Chicago, 5640 S Ellis Ave
AAC 010c, Chicago IL 60637}
 
\email{vikram@oddjob.uchicago.edu }

\begin{abstract}
Mass-loss from massive stars leads to the formation of circumstellar
wind-blown bubbles surrounding the star, bordered by a dense
shell. When the star ends its life in a supernova (SN) explosion, the
resulting shock wave will interact with this modified medium. In a
previous paper \citep{d05} we discussed the basic parameters of this
interaction with idealized models. In this paper we go a step further
and study the evolution of SNe in the wind blown bubble formed by a 35
$\msun$ star that starts off as an O star, goes through a red
supergiant phase, and ends its life as a Wolf-Rayet star. We model the
evolution of the circumstellar medium throughout its lifetime, and
then the expansion of the SN shock wave within this medium. Our
simulations clearly reveal fluctuations in density and pressure within
the surrounding medium, due to the changing mass-loss parameters over
the star's evolution. The SN shock interacting with these
fluctuations, and then with the dense shell surrounding the wind-blown
cavity, gives rise to a variety of transmitted and reflected shocks in
the wind bubble. The interactions between these various shocks and
discontinuities is examined, and its effects on the emission from the
remnant, especially in the X-ray regime, is noted. In this particular
case the shock wave is trapped in the dense shell for a large number
of doubling times, and the remnant size is restricted by the size of
the surrounding circumstellar bubble. Our multi-dimensional
simulations reveal the presence of several hydrodynamic
instabilities. They show that the turbulent interior, coupled with the
large fluctuations in density and pressure, gives rise to an extremely
corrugated SN shock wave. The shock shows considerable wrinkles as it
impacts the dense shell, and the impact occurs in a piecemeal fashion,
with some parts of the shock wave interacting with the shell before
the others. As each interaction is accompanied by an increase in the
X-ray and optical emission, different parts of the shell will
`light-up' at different times. The reflected shock that is formed upon
shell impact will comprise of several smaller shocks with different
velocities, and which are not necessarily moving radially inwards. The
non-spherical nature of the interaction means that it will occur over
a prolonged period of time, and the spherical symmetry of the initial
shock wave is completely destroyed by the end of the simulation.
\end{abstract}


\section{INTRODUCTION}

Core-collapse supernovae (SNe), those generally classified as Type
Ib/c and Type II, arise from massive stars (M $>$ 8 $\msun$). These
stars lose a considerable amount of mass prior to their explosion as
supernovae.  Mass loss from the star will modify the medium
surrounding it, giving rise to wind-blown cavities surrounded by
expanding shells of gas.  When the supernova explodes, the resulting
blastwave will eventually interact with this modified circumstellar
medium (CSM) rather than the interstellar medium (ISM) into which the
star was born.

In the past decades several pieces of evidence have suggested that
many supernovae arise within circumstellar bubbles. The most famous
example is the exceptionally well-studied SN 1987A, which is thought
to have exploded within a bipolar circumstellar bubble \citep{sck05a,
sck05b}. This bubble is due to the interaction of the wind from the
blue-supergiant progenitor with the mass loss from a prior red
supergiant (RSG) stage \citep{lm91, bl93}. Other well know SNe that
have been interpreted as arising within a wind-blown bubble include
Cas A \citep{bsbs96}, G292+0.8 \citep{pr02}, RCW86 \citep{vkb97}, and
the Cygnus Loop \citep{l97}.  Circumstellar interaction models have
been proposed for N132D \citep{h87} and Kepler's SNR \citep{b87}.

The formation of wind-blown bubbles around stars has been studied in
detail, both analytically \citep{wmc77, km92a, km92b} and numerically
\citep{fm94, mf95,dcb96}. Such models usually assume an ad-hoc
prescription for the wind properties.  Some attempts have been made to
take the evolution of the wind into account in the case of planetary
nebulae \citep{m95, db98} and Wolf-Rayet bubbles \citep{gm95, bd97}.

The evolution of supernovae in different media has also been studied,
especially in a constant density medium, and in a medium whose density
decreases as a power-law with radius. Self-similar solutions for the
interaction of supernovae with a power-law density profile have been
derived by \citet{c82}. Ostriker \& McKee (1988) present a compilation
of solutions for the evolution of astrophysical blast waves in various
media.

The density structure within a CS bubble however is quite different
from that in the constant density ISM. Thus the evolution of the
supernova remnant within the pre-existing wind-blown bubble differs
considerably from its counterpart evolving in the pristine ISM. The
evolution of supernovae in wind-blown bubbles has not received quite
as much attention as it deserves. \citet{cd89} did some preliminary
work, and \citet{cl89} studied the evolution analytically. A series of
papers by Tenorio-Tagle and his colleagues \citep{tbf90, trfb91,
rtfb93} studied the numerical evolution in further detail. A CS
evolution model for Cas A was computed by \citet{bsbs96}.

In \citet[][hereafter Paper 1]{d05} we introduced the various aspects
of the problem, and carried out one-dimensional calculations that
illustrated the various parameters involved. The results are
summarized in \S \ref{sec:intro}. These calculations assumed idealized
winds with constant properties, appropriate for studying the various
factors that affect SN evolution. More realistic models however
require a prescription of the mass loss history of the star as its
ascends the HR diagram, coupling the evolution of the circumstellar
gas to that of the star itself.

An important step forward in this process was the computation of
detailed stellar evolution models of 35 and 60 $\msun$ stars by
Norbert Langer \citep{lhl94}. These models, which provided the
required mass loss history with time, were used by \citet{glm96a,
glm96b} to compute the dynamical evolution of the surrounding gas. As
they show in their work, taking the complete stellar mass-loss history
into account results in a much more complicated circumstellar
structure.  The presence of a variety of small scale structures,
various dynamical instabilities and multiple shock fronts presents a
complex morphology that compares well with observations.  These
calculations however did not compute the full evolution of the
circumstellar bubble in multi-dimensions, nor the final supernova
phase of the star and the resultant interaction with the surrounding
medium.

In this paper we take a more detailed look at a specific case, that of
the 35 $\msun$ star. Using the 35 $\msun$ star model, courtesy of Prof
Norbert Langer, we first study the formation of the medium around the
star as it evolves along the HR diagram. We then assume that the star
explodes as a SN, leaving behind a compact remnant, and study the
evolution of the SN shock wave within this medium. Initially we
compute spherically symmetric one-dimensional calculations that
illustrate the various shock structures and the dynamics
involved. This is followed by two-dimensional computations that take
multi-dimensional factors such as deviations from symmetry, the onset
of turbulence, and the presence of hydrodynamic instabilities into
account. 

Our aim in this paper is to show the impact of a single massive star
on the surrounding medium throughout its lifetime. Although we
concentrate on a specific case, our goal is to illustrate general
properties of the interaction which are more globally applicable, and
to identify the various features which distinguish this from the
interaction of a SN shock with the ISM. Preliminary results from this
work were outlined by \citet{d07a}. This work expands considerably on
the results outlined therein, describes in detail the spherically
symmetric calculations that are necessary to understand the shock
dynamics and structures, computes X-ray luminosity and surface
brightness, and elaborates on the intricacies of the multi-dimensional
calculations. In a companion paper \citep{d07b} we also provide a more
general analytic discussion for the properties of wind-blown bubbles
around massive stars.

The rest of this paper proceeds as follows: In \S 2 we give a brief
overview of supernova explosions within wind-blown cavities. In \S 3
we present one-dimensional simulations that illustrate the evolution
of the bubble density and pressure with time, and describe the various
evolutionary phases. This is followed by a 1D simulation of the
SN-bubble interaction, which captures the essence of the
hydrodynamics. \S 4 follows up with 2D simulations that show the
formation of various dynamical instabilities and other
higher-dimensional effects. \S 5 summarizes the paper, provides a
general discussion of the results and their applications, and outlines
followup work.

\section{Overview of SNR-wind bubble interaction}
\label{sec:intro}

\citet{tbf90} and \citet{d05} showed that the interaction of the
supernova ejecta with a wind-blown shell can be divided into various
regimes, depending on the ratio of the mass of the shell to the mass
of the ejected material, a quantity that we label as $\Lambda$. If
this ratio is small ($ \Lambda << 1$) then the presence of the shell
merely acts as a perturbation to the flow. Indications of the
interaction are visible in the density, velocity and temperature
profiles of the ejecta.  However once the shock has swept up an amount
of material exceeding a few times the shell mass, the ejecta `forget'
about the existence of the dense shell. The ejecta density profile
reverts back to the profile that would have existed in the absence of
the shell.  The expansion parameter `$\delta$' (where R$_{sn} \propto
{\rm t}^{\delta}$) which had dropped considerably at the point of
shell interaction, increases gradually till it reaches the value it
would have had in the absence of the shell. In about 10-20 doubling
times, the SNR will completely "forget" about the shell and continue
to evolve as if the shell had never existed. The density profile
changes to reflect this. Since the emission from the remnant after a
few months is mainly due to circumstellar interaction, the changing
density profile will be reflected in a change in the emission from the
remnant, such as the optical and X-ray emission. Dwarkadas (2005)
showed the change in the X-ray surface brightness profile due to
changes in the density profile.

As the ratio $\Lambda$ increases, the energy imparted by the remnant
to the shell is larger, and the evolution begins to change. The
interaction of the remnant with the shell drives a shock front into
the shell. The high pressure behind the shock-shell interface sends a
reflected shock back through the ejecta. Thermalization of the ejecta
is achieved in a much shorter time as compared to thermalization by
the SN reverse shock.  The ejecta after reaching the center bounce
back, sending a weaker shock wave that will collide again with the
shell. In time a series of shock waves and rarefaction waves are seen
to be traversing the ejecta. Each time a shock wave collides with the
dense shell a corresponding (but successively weaker) rise in the
X-ray emission from the remnant is seen.

The presence of the dense shell results in a deceleration of the shock
wave, transfer of ejecta energy to the shell as well as conversion of
kinetic to thermal energy, which in the case of a very dense shell may
be effectively radiated away. These effects tend to speed up the
evolution of the nebula. The `Sedov stage' may be reached later than
for evolution in a constant density medium (due to the lower interior
density) and may last for a shorter time (due to the dense shell). In
some cases when the value of $\Lambda >> 1$ the Sedov stage may be
completely by-passed. In such a situation the shock wave will merge
with the wind driven shell. The velocity of the blast wave is
sufficiently decreased that the flow time becomes comparable to the
age of the remnant. Radiative cooling begins to dominate and the
remnant may enter the radiative stage much earlier on in its lifetime.

Herein we have outlined the basics of SN shock wave interaction with
dense shells.  Further details can be found in \citet{d05} and
references therein.

\section{Evolution of the CSM around a 35 Solar Mass star}

\subsection{1-dimensional calculations}

In this section we outline the evolution of the Wolf-Rayet bubble
around a 35 solar mass star, and describe the eventual SN-shell
interaction. Although this has been previous carried out by
\citet[hereafter GLM96]{glm96b} we chose to redo the entire
calculation. It is necessary for us to run the calculation in order to
be able to remap it onto the grid, and set up the initial conditions
for the SNR to to interact with this medium. In the process of
carrying out this simulation we found several differences between our
work and that of \citet{glm96b}, especially in the two-dimensional
case. Our use of a lower ambient density (more characteristic of the
ISM) allowed the bubble to expand to a much larger size. The
differences in our simulations are mainly due to us being able to
compute the evolution over all the stages in two-dimensions as a
result of the considerable increase in computational power. This is
reflected in a change in the character of the medium into which the
SNR evolves. Given the size and morphology variations in our
computations as compared to those of \citet{glm96b}, and the necessity
in describing accurately the medium into which the SN shock wave is
evolving, we have chosen to present herein a rather detailed
description of the evolution of the bubble.

The simulations were carried out using the VH-1 code, a
two-dimensional finite-difference hydrodynamic code based on the
Piecewise Parabolic Method \citep{cw84}. The code employs an expanding
grid, that tracks the outer shock front and expands along with
it. This trick is very useful in cases where the dimensions of the
grid expand by many orders of magnitude over the course of a run. The
1D simulations presented here were carried out on a grid of 2000
zones. 2-dimensional simulations were carried out on a grid of 600 by
600 zones. Radiative cooling was implemented in the form of a cooling
function, adopted from the one given in \citep{sd93}, and modified to
extend to lower temperatures. The effect of the lower temperature is
mainly to make the shells thinner.

In the course of its evolution the star evolves through 3 phases. The
wind properties in each phase are computed generally from
parameterizations derived from fits to the observed data
\citep{kppa89, l89, nj90}. The first and longest lasting is the
main-sequence (MS) stage which lasts for about 4.56 million years. In
this stage the star, which starts its life as an O star, loses mass in
the form of a wind with a mass-loss rate on the order 10$^{-7}\; {\rm
to}\; 10^{-6} \msun$~yr$^{-1}$ and a wind velocity that starts close
to 4000 km~s$^{-1}$ and gradually decreases with time. The wind
velocity of the star over its evolution is shown in Figure
\ref{fig:bubvel}, and the mass-loss rate is depicted in Figure
\ref{fig:bubmdot}. At the end of the MS phase the star swells up
immensely in size to become a red supergiant (RSG). In this phase mass
is lost in the form of a very slow, dense, wind. The velocity in this
simulation decreases down to 75 km~s$^{-1}$, although in reality we
would expect RSG winds to be even slower, on the order of 10-20
km~s$^{-1}$. The mass-loss rate increases to close to 10$^{-4}
\msun$~yr$^{-1}$. Since the RSG phase lasts for about 250,000 years,
the total mass lost in this stage is very large, about 19.6
$\msun$. At the end of the RSG phase, the star sheds its H envelope
and becomes a Wolf-Rayet star. Stars in this stage lose mass in the
form of radiatively driven winds, with a mass loss rate that is a
factor of a few smaller than their RSG predecessors, but a velocity
that is 2 orders of magnitude higher. As can be surmised, the
recurrent changes in the wind properties can lead to continuous
changes in the structure of the surrounding medium into which the
stellar wind is expanding.

In order to accurately compute the surroundings we assume that the
star was born in a medium with constant density 2.34 $\times$
10$^{-24}$ g cm$^{-3}$, a number density of about 1 particle/cm$^3$
for a medium with 90\% H and 10\% He. This density is lower than that
assumed by GLM96, who used an artificially high density to avoid large
computational domains. Our use of an expanding grid partially
circumvents the latter problem, since we do not need to start from a
grid extending out to about 100 parsecs. Our initial grid extends out
to about 10$^{15}$ cms. The lower external density results in a bubble
about twice the radius obtained by GLM96.

It is not clear what the density of the medium around the star is over
its lifetime. On the one hand many observations suggest that the mean
surface density of a massive star forming region is around 1 g
cm$^{-2}$ \citep[see review in][]{m04}, which would indicate a very high
volume density. But OB stars have high velocities that may cause them
to drift away from their birthplace \citep{mc05}. The presence of GRBs
occurring several hundred parsecs away from massive star forming
regions \citep{hfs06} shows that some massive stars do end their lives
in low density regions, irrespective of where they were
born. Observations of isolated HII regions which are far removed from
their host galaxies \citep{rw04} also indicate that sometimes massive
star formation can occur in very low density regions. Taking all these
factors into account, we have assumed a density of 1 particle/cm$^3$
as an average interstellar medium density over the stellar
lifetime. We note that if the density is higher, many of the results
will scale appropriately.

The interaction of the MS wind with the surrounding medium leads to
the formation of a double shocked structure, consisting of an outer
shock expanding into the ISM, and a reverse shock, often referred to
as a ``wind termination shock''. The termination shock separates the
free-streaming wind from the shocked wind region, and moves inwards in
a Lagrangian sense, i.e.~the entire structure expands outwards, but
the termination shock eventually moves towards the center with respect
to the outer shock. Most of the volume is occupied by the shocked wind
region, forming a hot, tenuous region that may emit in soft X-rays.
The formation of wind-blown bubbles is further described in Paper 1
and references therein. The density and pressure structure at various
timesteps during the evolution of the bubble in this stage are shown
in figure \ref{fig:ms1d}. The title of each plot gives the evolution
time in years. The two numbers in the top right hand corner denote the
wind velocity in km/s (upper) and the mass-loss rate in solar masses
per year (lower). The initial wind velocity is close to 4000 $\kms$,
and decreases slowly with time, accompanied by a corresponding
increase in its mass loss rate. This happens in such a manner that the
mechanical luminosity (0.5 $\times \dot{M} {v_w}^2$ ) is almost
constant. \citet{wmc77} calculated an analytic solution that describes
the evolution of the outer shock with time under such circumstances.
The radius R of the shell increases as R $\propto$ t$^{0.6}$, which
confirms reasonably well with the simulations throughout most of the
MS stage, and the structure of the bubble is reasonably consistent
with our expectations based on this paper. As mentioned the main
sequence stage in this model lasts for about 4.5 $\times$ 10$^6$
years. The total amount of mass lost in this stage is on the order of
2.5 $\msun$.

Towards the end of the main-sequence stage there is an abrupt drop in
wind velocity as the star swells up considerably in size, and enters
the red supergiant stage.  The large drop in velocity, and
corresponding rise in mass loss rate leads to a much higher density
for the RSG wind, and a change in the wind ram pressure. A new system
of pressure equilibrium is established in the bubble interior, and the
position of the termination shock adjusts accordingly (figure
\ref{fig:rsg1d}, top two panels).  It is not clear that a pressure
equilibrium can always be established. If the wind velocity is very
low, then the pressure at the base of the RSG wind will never be equal
to that in the MS shell. In this case, where the wind velocity is set
(in an ad-hoc fashion) to a minimum of about 70 km/s, the ram pressure
is sufficient to establish a new equilibrium (figure \ref{fig:rsg1d},
at time T $\sim$ 4.75e6 years). Once the pressure equilibrium reaches
steady-state a region of freely expanding RSG wind is seen, separated
from the surrounding medium by a shock. The RSG wind is abruptly
decelerated at this shock and piles up against it, forming a thin RSG
shell. Note that the RSG wind velocity is much smaller than that of
the material into which it is expanding. Thus the expansion of the RSG
wind into the MS wind does not result in a wind-blown bubble. Even
though the duration of the RSG stage is small (only about 230,000
years), the mass loss rate is 2 orders of magnitude higher than in the
previous stage. Thus the total amount of mass lost is estimated by
GLM96 to be about 18.6 $\msun$.

At the end of the RSG phase the star enters the Wolf-Rayet phase. The
wind velocity increases by over two orders of magnitude, to a value of
around 2000 $\kms$, while the mass loss rate drops by a factor of a
few. The fast, low-density wind from the W-R star collides with the
free-streaming RSG wind, creating a thin shell of swept up material,
and forming an inner shock that separates the unshocked and shocked
winds (figure \ref{fig:wr1d}, top left panel). Given the high velocity
of the W-R wind and the fact that the mass-loss is lower only by a
factor of a few, the momentum of the W-R shell considerably exceeds
that of the RSG shell. The W-R shell collides with the RSG shell in
about 10,000 years (figure \ref{fig:wr1d}, time T $\sim$ 4.79 million
years).  The collision leads to a reflected shock that moves back into
the unshocked W-R wind, and a transmitted shock that enters the thin
shell. The emergent shock wave drags the RSG shell along with it as it
enters the low density MS bubble.  Eventually this structure will
collide with the main-sequence shell (figure \ref{fig:wr1d}, time T
$\sim$ 4.88 million years). We note that this whole phase, which has
extremely important implications for the bubble and SN evolution, was
not modeled by \citet{glm96a}.  The energy transferred to the
main-sequence shell is not significant and does not cause appreciable
motion of the latter. The collision results in an increase in pressure
and a rise in the X-ray emission from the shell. The pressure behind
the compressed shell is sufficient to send a reflected shock expanding
back into the wind material, that ends up compressing the already
shocked material.  The wrinkles seen in the density profile are due to
small scale changes in the wind properties. This is partially due to
the fact that the boundary conditions are calculated in a
discontinuous fashion. It may be possible to interpolate over the
terminal wind velocity and mass loss rate as a function of time and
store them as continuous functions; however we have not attempted to
do so. Note that at the very end of the star's lifetime the wind speed
rises to about 3000 $\kms$, whereas the mass loss rate decreases
gently, resulting in a drop in the wind density. The reverse shock,
which is moving back into a high-density medium, suddenly finds itself
plowing through a much lower-density environment. The situation
finally reaches an equilibrium when the dynamic pressure of the freely
expanding wind equals the pressure behind the reverse shock.

The X-ray luminosity of the bubble over the evolution is shown in
Figure \ref{fig:xlumbub}. Our aim is to obtain a reasonable
approximation to the X-ray luminosity without carrying out complex
calculations, which is not the intention of this paper. We have
therefore used the approximate fit to the \citet{rcs76} cooling curve,
as suggested by \citet{cf94}: The cooling function $\Lambda = 2.5
\times 10^{-27}\, T^{0.5}$ ergs cm$^3$ s$^{-1}$ for temperature T $> 4
\times 10^7$ K, and $\Lambda = 6.2 \times 10^{-19}\, T^{-0.6}$ ergs
cm$^3$ s$^{-1}$ for temperatures 10$^6 < $ T $< 4 \times 10^7$ K.
This cooling function accounts for thermal bremsstrahlung at the
higher temperatures, and line emission in the lower range. We see that
throughout the early evolution of the bubble, the luminosity of the
bubble is quite low, a few times 10$^{32}$ ergs/s. The variations are
not as important as the average luminosity indicated.  Using the
calculators on the Chandra website we find that a bubble with this
luminosity in our galaxy at a distance of a few kiloparsecs would not
be observable by the Chandra X-ray satellite, as the background count
rate would exceed that from the source. The luminosity increases after
the appearance of the W-R wind, and especially after the collision of
the W-R wind with the RSG wind. The reason is that the density of the
RSG wind, into which the W-R wind is expanding, is much higher than
the density of the surrounding bubble. Thus the emission measure is
much larger for the X-ray emission from the W-R wind, than for that
from the earlier phases. As expected the emission decreases once the
W-R wind passes the RSG wind and encounters the low-density MS
interior. However it rises again once the expanding shock wave
collides with the dense MS shell.

We note that the X-ray luminosity is in general quite low. The
\citet{wmc77} picture suggests a very hot interior temperature, and
observable diffuse X-ray emission. However several authors have
pointed out the paucity of Wolf-Rayet bubbles with diffuse X-ray
emission \citep{cgg06, wcmw05,cgggw03}. Furthermore even the bubbles
seen show a much lower temperature than expected from the models. Our
results suggest that the emission measure from these bubbles is low
enough that diffuse emission may not be seen, and this may account for
some of the discrepancy. However, the temperatures in our models,
although lower than those suggested in the analytic models of
\citet{wmc77} due to the mixing and turbulence in the interior, are
still around a few times 10$^6$ to 10$^7$ K, much higher than those
actually observed. Perhaps this is partly due to the fact that much of
the observed X-ray emission may be coming from high-density,
lower-temperature clumps in the unstable, turbulent interior. We will
investigate this suggestion more carefully in future, using our
multi-dimensional simulations.

According to GLM96, the star at this point is in the transition phase
from a WN to a WC star. The model calculations terminate at this
point, and the density and pressure profiles at the end of its
evolution are shown in figure \ref{fig:bub1dfinal}. Note that the
overall structure is quite similar to what one would expect from a
two-wind interaction, although there are considerable fluctuations in
the density profile in the bubble interior. In particular there is one
region in the bubble interior where there is an increase in density by
almost two orders of magnitude compared to the surroundings. This will
be somewhat smoothed out by instabilities in multi-dimensions.

It must be kept in mind that the evolution described above is only
1-dimensional.  In two dimensions, as we shall show below (\S
\ref{sec:bub2d}), strong instabilities may develop in the shells, the
interior of any surrounding bubble will not be isobaric, and the
structure of the circumstellar medium is not as clearcut as in the ID
models. It is however clear from this picture (Figure
\ref{fig:bub1dfinal}) that in general the medium surrounding a massive
star is a low-density medium, surrounded by a high-density thin
shell. This is similar to what was assumed in the calculations carried
out in Paper 1. It is significant also that the shell size is
essentially set by the MS stage, while the composition of the bubble
consists mostly of material emitted in the RSG phase. This material
would not have gone too far however were it not for the high-momentum
W-R wind, which mixes all the material out to the radius of the dense
shell. Thus what we refer to as a ``Wolf-Rayet bubble'' is the
cumulative consequence of the previous stages of evolution.

\subsection{Interaction of the SN shock with the Surrounding Medium}

\label{sec:sncsm1d}

The 35 $\msun$ model presented here ends its evolution as a 9.15
$\msun$ star. In order to study the further evolution, we assume that
the star subsequently explodes as a supernova of energy 10$^{51}$
ergs.  The outer layers are expelled in the explosion, leaving behind
a remnant neutron star of mass 1.4 $\msun$. Thus the amount
 of mass ejected in the explosion is about 7.75 $\msun$. Given the
 mass and energy, one more parameter, the exact form of the ejecta
 density profile, is needed in order to model the SN explosion. As
 described in Paper 1, we have assumed that the ejecta are well
 described by a power law profile in the outer parts, with a power-law
 index of 7, and a constant density profile in the interior
 \citep{cf94}. The choice of power-law index is necessarily arbitrary
 and meant to be illustrative, although it is useful to remember that
 less steep power-laws are thought to be more appropriate to describe
 the ejecta of SNRs arising from compact stars \citep{bsbs96}.

The initial interaction of the supernova shock with the free wind was
carried out separately on an expanding grid. The wind parameters used
were those existing at the end of the star's W-R stage. The shock
structure was then interpolated and mapped back onto the first few
(typically 75-100) zones of the grid containing the final stages of
the W-R star. Inflow boundary conditions are used.

The innermost wind termination shock of the W-R bubble is at a radius
of about 11.2 pc.  The SN shock takes about 880 years to reach this
point. Note that this time is shorter than that predicted by the
self-similar solution \citep{c82, cf94}. The reason is that the
self-similar solution assumes the external medium has a negligible
velocity, whereas in this case the ambient velocity (due to the W-R
wind) is a finite (and non-negligible) fraction of the SN shock
velocity. The collision of the SN shock with the inner wind
termination shock results in a reflected shock moving back into the SN
ejecta and a transmitted shock advancing into the low-density
bubble. The reflected shock can be seen (Figure \ref{fig:snbuba1d},
3743 years ) climbing the steep power-law part of the ejecta profile.

The SN shock wave at the start of the simulation was moving close to
13,000 $\kms$ (see Figure \ref{fig:snvel}). The collision with the
termination shock reduces the velocity considerably, and the
transmitted shock emerges into the interior cavity of the MS bubble
with a velocity close to 5,000 $\kms$. The shock continues to sweep up
more of the surrounding material, but the low density implies that the
swept-up mass is quite small, and therefore its velocity does not
change appreciably. As shown in Figure \ref{fig:snbuba1d}, a
high-density perturbation exists about 41 pc from the center, with a
maximum density of about 0.01 particles cm$^{-3}$ about 47 pc from the
center. The collision of the transmitted shock with this region
results in a corresponding drop in velocity, and a weak reflected
shock whose presence is mainly discernible as a slight perturbation in
the pressure profile, but which tends to otherwise blend in the
density structure (Figure \ref{fig:snbuba1d}, 9299 years). When the
shock hits the highest density part of this region (at about 47 pc)
another reflected shock is seen which is slightly stronger (labeled
$r_1$ in \ref{fig:snbuba1d}). The collision compresses the dense
region, increasing its density (Figure \ref{fig:snbuba1d}, 12531
years), and results in a transient increase in the X-ray emission
({figure \ref{fig:xlumsn}).

The transmitted shock that emerges from this collision has a velocity
that is closer to 2000 $\kms$. Note that this is lower than the
velocity the shock would have if it were expanding in a medium with a
constant, low density equal to the bubble interior density. It is the
frequent collisions with the higher density fluctuations that tend to
slow it down. The forward shock moves roughly at constant velocity
till it collides with the main sequence shell, at about 22000 years
(Figure \ref{fig:snbubb1d}). The mass of the shell far exceeds the
mass of the shocked material colliding with it, so this collision
falls in the regime where $\Lambda >> 1$ (see Paper 1).  The shock
front merges with the shell, and its velocity drops almost to
zero. Since the shock has essentially merged with the shell it is very
difficult to track the shock and note its actual radius, and hence
velocity, resulting in the velocity fluctuations seen after about
23000 years in Figure \ref{fig:snvel}. Meanwhile, a strong reflected
shock resulting from the collision with the dense shell moves back
into the already shocked material.  This reflected shock collides with
the remains of the previously shocked high-density perturbation,
resulting in a weaker forward shock that expands outwards and
subsequently impacts the main-sequence shell (Figure
\ref{fig:snbubb1d}, around 30000 years). The result of all these
frequent collisions is a plethora of shock waves criss-crossing the
region of the remnant in the vicinity of the main-sequence shell. A
high-pressure region is formed there, and some weaker shocks are later
seen to expand outwards and subsequently collide with the
main-sequence shell (Figures \ref{fig:snbubc1d} and
\ref{fig:snbubd1d}).

In the meanwhile, the reflected shock $r_1$ from the earlier
collisions has reached the flat part of the density profile. This
reflected shock moves towards the center at a rapidly increasing
velocity, with maximum velocities approaching 6,000 $\kms$. On
reaching the center, and given our inflow boundary conditions, the
shock bounces back. This much weaker re-reflected shock will expand
outwards and eventually collide with the main-sequence shell in about
another 30,000 years, and the cycle tends to repeat itself (Figures
\ref{fig:snbubc1d} and \ref{fig:snbubd1d}). However, the SN density at
the origin (center) is decreasing as t$^{-3}$. The density in the
interior increases when the ejecta are shocked, although it will then
again decline with time. In general then the shocks are continuously
interacting with material of lower density, and therefore any
radiation signatures from the interaction, such as increasing X-ray
emission, are continually diminished. In fact the subsequent reverse
shock interactions with the shell hardly result in an increase in the
X-ray luminosity of the remnant, although the interaction of the
re-reflected shock with the dense main-sequence shell certainly leads
to a noticeable increase.

A few salient points of the interaction are noticeable in Figures
\ref{fig:snbuba1d} to \ref{fig:snbubd1d}:

\begin{enumerate}
\item At several times during the evolution, a variety of reflected,
transmitted and re-reflected shocks are visible in the SN density, and
especially pressure profile (e.g.~at time 89485 years and 100187
years). The pressure and density profiles are very different from
those assumed for a SN interacting with a constant density medium or a
wind, and therefore the emission computed from these will vary.
\item In the 1D calculations the high-density fluctuation that is
initially present is visible almost throughout the simulation. In
multi-dimensions this will probably be somewhat flattened due to the
presence of instabilities. But fluctuations in density will exist in
multi-dimensions, and the simulations reasonably depict the behavior
in such cases.
\item Once the SNR shock impacts the dense shell the evolution is more
or less restricted to the W-R bubble for 5-6 doubling times, and the
motion of the shell is negligible. Thus the size of the remnant in
this period is confined to that of the W-R bubble, and the remnant
will appear to have stalled.
\item Changes in the radiation signatures during this time are almost
entirely due to the effects of the reflected and other shocks within
the bubble, and the forward shock has very little role to play.
\item The complex surrounding structure results in a variety of shock
waves traversing the bubble at any given time. A large range of gas
velocities will be observed over the interior of the remnant. When the
SNR shock is heading towards the dense shell, gas velocities ranging
from -2000 km/s to +2000 km/s are seen in the interior. Once the shock
hits the shell the forward shock velocity is considerably reduced, but
the gas velocity behind the reverse shock can increase to 5000 km
s$^{-1}$ as the shock expands in a continually lower density medium.
Thus line emission or absorption spectra from different parts of the
remnant will show vastly different velocities, sometimes differing by
thousands of km s$^{-1}$. If a spectrum is taken that shows lines
arising from different parts of the remnant, it will reveal a very
confusing and complicated velocity structure.
\item By the end of the calculation the transmitted shock can be seen
expanding outwards, pushing the dense shell with it. It will take
considerably more time before this shock separates from the shell and
is visible as a separate entity (see Paper 1).
\end{enumerate}

The evolution of the SNR is thus very different from that seen if it
had expanded in the pure interstellar medium. The shock wave entering
the MS shell causes the shell to expand. As shown in paper 1, the
shock more or less merges with the shell in this case, and the
transmitted shock does not seem to appreciably separate from the
shell. Over time the density of the shell decreases, but its thickness
appears to increase as the forward shock expands without effectively
separating from the shell.  Even as late as 150,000 years, or over 6
doubling times, the outer shock is seen at a distance of about 84 pc,
which means that is has moved just 3 pc in about 125,000 years. A
large fraction of the kinetic energy has been converted to thermal
energy and emitted as radiation. The ejecta are completely thermalized
even before reaching the so-called Sedov stage. In fact the remnant at
this stage does not fit anywhere in the classical evolutionary pattern
of free expansion, Sedov, radiative stage and eventual dispersal into
the ISM. It comes closest to being somewhere in between the Sedov and
radiative stages, much closer to the latter.

An important point to note is that most of the gas within the bubble
is shocked, and is at temperatures 10$^6$ degrees or higher.
Occasionally a passing shock will raise the temperature at some point
to greater than 10$^7$ K. The entire remnant will appear luminous in
X-rays, although the emission from the outer parts dominates through
much of the evolution. The inner regions may brighten up when the
reflected shocks collide with them. As the X-ray surface brightness
depends on the square of the density, we have computed this value at
various different times (Figure \ref{fig:xsbsn}, for details see
Paper 1). The quantity plotted in the various frames is the square of
the particle density, integrated along the line of sight, at all
points where the temperature exceeds 10$^6$ K, and normalized to the
maximum value.  This quantity shows approximately the surface
brightness of the remnant in X-rays. As can be seen, in most cases the
outer regions appear to dominate the surface brightness. It is
noteworthy that in the last few plots in Figure \ref{fig:xsbsn}, the
emission appears to arise from the entire remnant, but the emission
measure at this point is so low that it would likely not be
observable.

The total X-ray luminosity from the remnant varies considerably with
time, depending strongly on the behavior of the various shock waves
criss-crossing the remnant. While an exact computation of the X-ray
emission, which would involve taking non-equilibrium processes into
account during the early evolution, is far beyond the scope of this
paper, Figure \ref{fig:xlumsn} illustrates the evolution of the X-ray
emission over the first 150,000 years of the remnant, in a similar
manner to Figure \ref{fig:xlumbub}.  The X-ray luminosity is on
average about 10$^{34}$ ergs s$^{-1}$, with occasional periods when it
increases by a few orders of magnitude.  A large extended source with
this average luminosity at a distance of 10 kpc would hardly be
visible with the Chandra satellite above the galactic background. Thus
the presence of the low density CSM considerably reduces the emission
from the remnant. However the periodic brightening would increase the
visibility considerably. The initial increase in luminosity is due to
the collision of the shock wave with the MS shell, while the secondary
maximum at about 89000 years is due to a combination of the
re-reflected shock hitting the main sequence shell and various
reflected shocks moving back into the ejecta. This shows that even
older remnants may experience some brightening in X-rays if the
supernova explodes within a pre-existing cavity, rendering them
visible even at late times.

The above description illustrates the differences between the
evolution of SNRs in a constant density medium as compared to more
realistic structured environments sculpted by the pre-supernova
progenitor star. In this section we have shown 1D calculations that
accurately track the expansion of the SN shock wave through the
surrounding medium. The 1D simulations can capture the dynamics of the
multitude of shocks that appear to criss-cross the remnant at various
times, and enable us to understand the complicated hydrodynamics and
kinematics. However, in order to obtain a complete picture one needs
to carry out multi-dimensional simulations that can capture effects
such the formation of hydrodynamical instabilities and deviations from
spherical symmetry which cannot be seen in one-dimensional
simulations. To this effect we have also carried out two-dimensional
simulations, as described below.

\section{Two-Dimensional Computations}

\subsection{Evolution of the surrounding Medium}
\label{sec:bub2d}

Unlike GLM96, we have chosen to carry out 2-dimensional simulations
from the start of the MS stage. While computationally this is
extremely time-consuming, since the simulation runs for over a million
timesteps, we feel that a thorough treatment of the problem requires
2-dimensional modeling of the entire evolution from the beginning of
the star's life. As we will show below, the MS bubble in two
dimensions shows a much more complicated structure, with a
non-isobaric interior that shows considerable density and pressure
fluctuations. In our simulations, the RSG and WR shells also become
unstable. Only the instability of the WR shell was noticed by
GLM96. Our simulations show differences from theirs right from the
start. Besides, they did not model the interaction of the WR wind with
the MS shell and its consequences, which are important to us to define
the pre-SN stage.

The two-dimensional simulations were carried out on a spherical
($r-\theta$) grid. The simulations described used 600 zones in both
the radial and azimuthal directions. In order to accurately compute
the inner boundary conditions, i.e. the velocity, the density from the
mass-loss rate and wind velocity, and the temperature, the simulation
runs for about 1.79 million time steps, one good reason why it was not
carried out in full two-dimensions by GLM96.

In figure {\ref{fig:bub2d_ms}} we show density images of the evolution
of the MS bubble at various timesteps. The radius scale is in
parsecs. In the top right corner of each figure we show the current
wind parameters - the velocity in km s$^{-1}$ and the mass-loss rate
in $\msun yr^{-1}$ - as well as the time in years. The color scale
shows the logarithm of the gas density, calibrated in g cm$^{-3}$.

The evolution of the MS bubble in two dimensions starts off as in the
1D case. A thin shell of swept-up material is formed, bounded on the
outside by a highly radiative shock and on the inside by a contact
discontinuity. The outer shock sweeps up the surrounding ISM into a
thin shell. The expansion of the bubble closely resembles that shown
in the previous sections and in \citet{wmc77}, with the radius
increasing approximately as t$^{0.6}$. The interior is initially
isobaric. However after about 75,000 years minor perturbations in
density and pressure seem to appear within the interior. Perturbations
appear to start close in to the inner shock, and somewhat later just
inside of the contact discontinuity. We have not added any
perturbations to the initial conditions. The origin of the
perturbations can be traced to irregularities that arise at the inner
shock, and we suggest that it is the response of the inner shock to
the changing wind parameters that gives rise to the fluctuations.  As
the wind parameters vary, the shock position varies correspondingly at
every timestep. The variation in the position of the inner shock
causes variations in the pressure around the shock, and since the
shock position is continuously varying, and the interior is subsonic
with respect to the reverse shock, the pressure waves do not have
enough time to isotropize within the interior. This leads to further
variations and density inhomogeneities, which result in the
development of turbulence in the interior, visible in Figure
\ref{fig:bub2d_ms}.

The fluctuations in the position of the shock front, and the shear
associated with it, result in a considerable amount of vorticity being
deposited into the shocked wind. As the shocked wind is expanding
outwards, the deposited vorticity is carried out with the wind, and
does not dissipate to smaller scales. Pressure variations due to the
deposition of vorticity near the inner shock lead to density
variations within the region, as regions at different temperatures
cool differently. The net result is the formation of higher density
regions within the interior. The stellar wind is forced to flow around
these obstacles, leading to slower moving regions within the radial
flow. The buildup of vorticity in these regions is clearly
demonstrated in Fig {\ref{fig:bub2d_ms}}.  The cumulative effect is
the growth of turbulence within the interior.

At the same time, we find that the dense, thin shell also shows some
signs of shear instabilities. Although we do not include any initial
perturbations in our calculation, small perturbations can be initiated
within the shell due to the shear flow between the contact
discontinuity and the cavity just interior to it, due to the
difference in the flow velocities. Such perturbations can already be
seen in the top right panel in figure {\ref{fig:bub2d_ms}}. These
instabilities tend to persist throughout the growth of the bubble, but
their effect is minor, and does not lead to any significant distortion
of the spherical shell. This is clearly seen in the last panel (bottom
right), which shows the structure of the bubble towards the end of the
MS stage. This complicated turbulent structure is the medium into
which the RSG wind will expand.

We note here that in previous low-resolution simulations \citep{d01,
d02, d04}, we have reported the presence of an unstable thin shell in
the main sequence phase, which we attributed to a Vishniac-type thin
shell instability \citep{v83, vr89}. This was seen in simulations with
grid resolutions from 200 $\times$ 200 to 400 $\times$ 400. However,
this instability disappeared when we ran simulations with the even
higher resolution reported herein. We feel that the higher resolution
simulations are far more believable, although we have been unable to
understand the reasons for triggering of the instability at lower
resolutions. However, the presence of this instability (or lack of it)
does not affect the subsequent evolution of the bubble or the SN
shock, which is our main concern for this paper. The remaining
behavior is seen as expected at both lower and higher resolutions,
with the higher resolutions providing increasingly sharper clarity to
view the R-T instabilities and turbulence within the interior. It is
curious though that we have noted similar behavior when carrying out
simulations of a 40 $\msun$ star from a stellar model provided by Dr
Georges Meynet. The MS shell instability was seen at a resolution of
400 $\times$ 400 zones, but not at a higher resolution of 600 $\times$
600 zones. The current situation should be unstable to the Vishniac
ram pressure instability \citep{v83}, but the appearance of this would
differ from the finger-like perturbations that we had written about
earlier. Thus the presence (or absence) of an instability in the MS
shell is still a puzzle that remains to be solved.

The interior of the cavity thus differs considerably from the
isotropic spherical bubble assumed by GLM96.  At the end of about 4.5
million years, with the MS shell at a radius close to 75pc from the
star, the wind velocity begins to decrease rapidly as the star enters
the RSG stage. As mentioned earlier, the higher mass loss rate and
lower velocity of the RSG wind causes the pressure equilibrium within
the bubble to change, and a termination shock is formed where the ram
pressure of the RSG wind is equal to the thermal pressure in the
interior. The RSG shell is also found to be unstable, as shown in
figure \ref{fig:bub2d_pms}. GLM96 suggest that the shell is stable to
Rayleigh-Taylor instabilities, but this would be strictly true only if
the wind velocity was constant or monotonic.  In this case the RSG
wind velocity varies from about 95 $\kms$ to about 70 $\kms$ over the
evolution, and the pressure equilibrium also varies. The RSG shell is
found to be decelerating as it expands outwards. The high density
behind the shock decelerated by the high pressure of the wind-blown
cavity within the MS bubble provides the right conditions for the R-T
instability, and we see clearly the growth of Rayleigh-Taylor fingers
in our simulations. As the dense shell is being decelerated the
fingers tend to expand outwards. Figure \ref{fig:bub2dzoom} zooms in
on the RSG region towards the end of the RSG stage, showing vividly
the thin expanding R-T filaments, with slightly bulbous heads due to
the formation of Kelvin-Helmholtz instabilities at the tip of the
fingers, resulting from the shear flow between the fingers and the
surrounding material. The filaments grow in length over the evolution
of the shell.

At the end of the RSG stage the surface temperature of the star begins
to increase, it sheds its outer envelope and enters the W-R phase. The
WR wind expanding into the RSG wind will form a miniature version of
the MS bubble, complete with an inner shock, a contact discontinuity
and a thin shell. The W-R shell is also found to be unstable and
filamentary structure can be seen at the inner edge of the shell. We
attribute the growth of these structures also to the development of
the Rayleigh-Taylor instability.

In the ideal course of the collision of two winds with constant
wind-properties, the wind-blown shell would expand with a constant
velocity, and the conditions would not allow for the development of
the R-T instability. However in this particular case, although the W-R
wind is expanding within the RSG wind, there are some important
differences. Firstly the wind parameters are not constant, either in
the W-R or the RSG stage. Secondly, the W-R bubble is so highly
supersonic that it does not cool efficiently, and therefore the outer
shock is not highly radiative, and the shell that forms is not
thin. The formation of a thin, dense, cool shell may have led to the
development of thin-shell instabilities. Instead due to the variation
in parameters we find that the W-R shell is accelerating down the ramp
of the RSG wind. The pressure inside the dense shell (in the interior
cavity of the W-R bubble) is much larger than the pressure outside (in
the RSG wind). The high pressure pushing on the high density shell
provides the appropriate conditions for the R-T instability to
develop. In this case however, as opposed to the RSG case, the fingers
expand inwards, from the high density shell into the bubble, rather
than outwards, as is expected from the physical conditions. The
structure at this stage is shown in Figure \ref{fig:bub2d_pms}, and
magnified in the upper right panel in Figure \ref{fig:bub2dzoom}.

It is interesting to note that although the formation of the R-T
instability is seen in both the RSG and W-R cases, the structure looks
quite different. One reason is that in the RSG case the shell is
decelerating outwards, whereas in the W-R case the shell is
accelerating outwards. Another reason is that the density contrast, or
the Atwood number, is very different between the two cases. The time
that the instability can develop is very short in the W-R case, as the
W-R wind quickly advances in the RSG wind before colliding with the
shell. For these reasons the appearance of the R-T fingers is
different in the two cases, as is apparent in figure
\ref{fig:bub2dzoom}. In our simulation we do not have enough
resolution in the radial direction to study the instability in the W-R
wind in great detail.

The extremely large velocity of the W-R wind (on the order of 3300
$\kms$), coupled with a mass loss rate which is only a factor of a few
smaller than that of the RSG, means that the ram pressure of the W-R
wind far exceeds that of the RSG wind.  The W-R wind slams into the
RSG shell, causing it to fragment completely, and is seen to break-out
of the shell at many different points (Figure \ref{fig:bub2dzoom}),
primarily along the axis, (although this may be a numerical
effect). The formation of the funnel-like feature close to the axis is
certainly a numerical effect, due to the presence of very narrow zones
along the axis. However the simulation clearly reveals the break-up of
the RSG shell, with the fragments being dispersed throughout the
interior of the W-R bubble.  The momentum of the W-R wind carries some
of the RSG material along with it towards the MS shell (Figure
\ref{fig:bub2d_pms}).  The W-R gas collides with the MS shell and
bounces back, in the process distributing the RSG material throughout
the interior of the nebula. Thus although the RSG wind itself was not
able to penetrate even a fifth of the MS radius, much of the mass in
the interior of the nebula will be comprised of material dredged up
from the RSG stage. This material is then shocked by the W-R wind,
perhaps leading to an overabundance of N and/or C in the nebular
material. The collision of the W-R wind with the MS shell results in a
shock being driven into the shell and a reflected shock back into the
unshocked wind. This reflected shock penetrates as far back as it can
towards the center, before its advance is arrested by the ram pressure
of the outflowing W-R wind, and an inward facing wind-termination
shock forms.

Figure \ref{fig:bub2dzoom}, bottom right panel, shows a combined
contour and vector plot of the structure of the bubble at the end of
the simulation. The complex nature of the velocity field is due to the
various evolutionary phases, although the imprint of the W-R phase is
unmistakable in the inward flow near the axis and the complicated
behavior in the equatorial regions. The intersection of the ram
pressure of the radially outward flow and the thermal pressure behind
the reflected shock leads to the formation of the W-R wind termination
shock. It is of consequence to note that the wind termination shock is
{\em not spherical} but slightly elongated towards the equatorial
latitudes. This is due to a combination of factors. The unstable W-R
wind pushes out on the unstable RSG wind, fragmenting the RSG
shell. The material is not carried out in a spherically symmetrical
manner. This asphericity is enhanced when it travels through the
cavity due to the pressure and and density fluctuations, and again on
this material colliding with the shell. Therefore, when the reflected
shock's progress towards the center is finally halted, the equilibrium
between the isotropic pressure of the W-R wind and the varying thermal
pressure behind the reflected shock leads to an aspherical wind
termination shock. This has important consequences for the subsequent
evolution of the SN shock wave.

Qualitatively a radial cut through the nebula resembles the 1D
profile. However there is considerably more structure in the 2D
profile.  Also since the 2D resolution is about a third of the 1D case
the structure is more smeared out, and the sharp shock fronts of the
1D model are spread out over a larger distance in 2D.

We note that our final picture of the W-R bubble agrees well with
observations, including the large size and the complicated internal
structure \citep{cec03}. The size of the bubble of course is a direct
consequence of the external density that we have assumed. A much
higher density would lead to a smaller size. This would however just
compress the entire picture into a small radius, and therefore
increase the density in the bubble interior, but it would not
appreciably change the dynamics and kinematics that we see.

\subsection{Evolution of the SN Shock Wave}

In the next stage the star was assumed to undergo a supernova
explosion as outlined in \S \ref{sec:sncsm1d}. The SN profile was
interpolated onto the CSM grid. The SN profile occupies 63 zones, and
the interpolation results in a smearing out of the shock front and
other features. The simulation was then run to study the evolution of
the SN shock wave into the surrounding medium. In this discussion we
will concentrate on describing the main multi-dimensional effects and
departures from the spherically symmetric case discussed in \S
\ref{sec:sncsm1d}. 

The overall evolution of the SN shock wave proceeds as in the 1D case,
but with one major difference - the SN shock does not remain
spherical. We elaborate on this aspect, and its consequences, below.

The evolution of the SN shock wave in the freely expanding wind
proceeds as expected, and in a fashion similar to the 1D case, up
until the time that the shock reaches the wind termination shock. Note
that due to the aspherical nature of the wind-termination shock, the
interaction first takes place in the region of the symmetry axis. This
is shown in Figure \ref{fig:snbub2dpre} (at 2597 years), which shows
snapshots of the pressure at various times during the evolution. The
pressure is chosen as a quantity to highlight the shocked interaction
region between the forward and reverse SN shocks. As explained in \S
\ref{sec:sncsm1d}, the interaction results in a transmitted and a
reflected shock wave. The transmitted shock wave has been decelerated
by the interaction with the wind shock, and is therefore slower than
the rest of the shock, which is still undecelerated as it has not yet
encountered the wind shock. Thus this portion lags slightly behind the
rest of the shock wave. As the next part of the shock wave hits the
wind shock, it also gets decelerated. Since the velocity of the SN
shock is decreasing as it moves outwards, each subsequent impact of
some part of the SN shock with the wind shock happens at a lower
velocity. The net result of this impact is that different parts of the
transmitted (and reflected) shocks travel outwards at slightly
different velocities. Both shock waves assume the shape of the
aspherical wind shock to a certain degree.

The effect is further accentuated by the fact that the interior cavity
is not isotropic, but shows considerable variation in pressure and
density, with several regions that are higher density than the
surroundings. Furthermore, as shown in Figure \ref{fig:bub2dzoom},
lower right panel, the interior also contains several vortices with
velocities as high as a thousand km/s. The interaction of the SN shock
wave with this highly turbulent material, and the fact that the
average pressure behind the shock wave is not significantly higher
than the pressure of the interior, causes further corrugations in the
shock wave. The result is a highly wrinkled shock wave, with various
bumps and wiggles, by the time it reaches the outer dense shell.

Figure \ref{fig:snbub2dden} illustrates how the spherical shock slowly
becomes a wrinkled and corrugated structure. In order to display this
effect vividly, we plot the density profile of the shockwave at
various timesteps as it expands towards the wind-blown shell. Although
the outer and inner shock are not very well resolved in the density
color scale, it is easy to spot their location, especially using
figure \ref{fig:snbub2dpre}. In Figure \ref{fig:snbub2dden} at time
4932 years, the shock wave is encountering a density fluctuation
within the bubble. The shock appears slightly depressed in that
region, and a reverse shock (purple in color) can be seen reflecting
off the perturbation. At later times this event is repeated, adding to
the asymmetry of the shock wave each time, until finally the shock
wave becomes extremely distorted just before it is about to collide
with the shell, at about 20000 years.

Since the shock is not spherical, the expansion is not completely
radial. The bumps in the shock wave, and the crinkled nature of the
shock, results in just one or two extended sections of the shock
hitting the dense shell at about 22,000 years, as opposed to the
entire shock wave in the one-dimensional case. Each collision will
result in a rise in the X-ray and optical emission, as we have seen
before (see Paper 1). However, in this case, since the shock collides
with the shell in a piecemeal fashion, different parts of the shell
will brighten up in the X-ray and optical at different
times. Therefore, instead of seeing a glowing shell, what will be seen
are different sections of the shell ``lighting up'' at different
times. Eventually, in this case over a timescale of about 15000 years,
the entire shell will brighten up as the entire extent of the shock
has collided with the shell.

Another effect of this piecemeal collision is that instead of having
one reflected shock bouncing off the shell, we will have several small
``shocklets'', with velocity vectors pointing in different
directions. This results in an even more aspherical reverse shock, as
is shown in figure \ref{fig:snbub2dpre} after about 35000
years. Different parts of the reverse shock will then move at
different velocities towards the center. Since the velocities are not
all radial, some portions will advance preferentially towards the
symmetry axis or the equatorial axis, and collide with it. In the last
panel in figure \ref{fig:bub2dzoom}, just after 45000 years, one part
of the reflected shock can be clearly seen to have collided with the
axis of symmetry, and a re-reflected is just starting to move
back. Note that this shock is directed almost perpendicular to the
axis. In part this is due to an axis effect, as has often been
discussed for two-dimensional axisymmetric simulations. Therefore we
caution into reading too much into the {\em specific} behavior of the
shock outlined in this case, emphasizing more the existence of an
overall global asymmetry, and reflected shocks that are not moving
radially inwards.

Due to this piecewise interaction the reflected shock takes a longer
time to reach the center in multi-dimensions. In the one-dimensional
case the reflected shock takes about 31000 years or so to reach the
center and bounce back. In the two-dimensional case, only a very small
fraction of the reverse shock has reached the center in almost 45,000
years. This is not surprising, considering that the entire forward
shock itself has barely just finished colliding with the dense shell,
and that the reverse ``shocklets'' that are formed have a significant
azimuthal component rather than just a radial, centrally directed
component.

The SN shock wave gets essentially trapped in the dense shell in this
case, as outlined in \S \ref{sec:sncsm1d}. A transmitted shock
eventually emerges, as shown in the one-dimensional case. However in
this case the transmitted shock takes an even longer time to emerge as
compared to the one-dimensional case, for the same reasons that we
have outlined above, mainly the slower motion of the SN shock wave and
the larger time taken for the entire shock wave to collide with the
shell. This will thus introduce an even larger degree of asymmetry in
the transmitted shock wave as different parts of the shock emerge from
the shell at different times. The appearance will be of a very
aspherical remnant. Unfortunately due to computational time
constraints we have not carried out the simulations further. In future
we plan to use a parallel adaptive mesh code to carry out this
simulation in three-dimensions, thus removing any axis of symmetry.

The final panel in Figure \ref{fig:snbub2dden}, at about 42000 years,
displays the density profile long after shock-shell interaction. The
filamentary nature of the resultant remnant is very clearly evident in
this picture, resulting from dense shell material expanding into the
low density cavity after the shock-shell interaction, suggestive of
Richtmeyer-Meshkov and Rayleigh-Taylor instabilities. The final
morphology is a combination of the various asymmetries and the
piecemeal collision, together with the various pieces of reverse shock
that also interact with each other. The result is a spaghetti-like
mesh of filaments emanating from the inner walls of the nebula. The
color scale gives the logarithm of the density. It can be seen that
the filaments are about two orders of magnitude higher in density.

It is clear from the description that both the transmitted and
reflected shocks may be considerably aspherical. Furthermore, the
radial symmetry of the remnant is gradually lost over increasing
interactions. In our case we started with the SN shock evolving in a
highly turbulent, but still spherical bubble, yet ended up with a
structure with a very aspherical shock wave and reflected shock. In a
case where the circumstellar bubble around the remnant is itself not
spherical but bipolar, such as the very well observed SN 1987A, this
effect will be even more pronounced. Thus many of our simplistic
notions of radially expanding outgoing and incoming shock waves in
SNRs need to be re-evaluated.

\section{Summary and Discussion}

Continuing our series of papers on the evolution of SNe in structured
wind environments, we have herein explored the case of a 35 $\msun$
Wolf-Rayet star via numerical simulations. The star starts its life on
the main-sequence as an O star, evolved through the red supergiant
phase, and then becomes a Wolf-Rayet star. As it evolves, it loses
mass via winds, whose properties change dramatically over the entire
evolution. The mass-loss leads to the formation of a structured
wind-blown bubble around the star. A main-sequence bubble with a dense
shell is formed initially. The slow and dense RSG wind does not expand
very far inside this shell, and its low velocity does not result in
the formation of a wind-blown bubble, but a termination shock is
formed and the wind piles up against it. The high-momentum of the
Wolf-Rayet wind pushes the RSG material outwards, until it collides
with the MS shell and rebounds back. Finally a Wolf-Rayet wind
termination shock is formed at the radius where the ram pressure of
the freely expanding wind and the thermal pressure behind the shock are
equal.

This is the basic description of the evolution. Multi-dimensional
calculations add further details to the overall picture. The
constantly fluctuating position of the reverse shock in the MS phase
results in the buildup of pressure fluctuations and the deposition of
vorticity into the shocked wind. The vorticity is carried out with the
shocked expanding flow. These effects result in the formation of
eddies and the onset of turbulence within the shocked medium.

The RSG wind and the W-R wind shell are also found to be unstable to
R-T instability. The high-momentum W-R wind pushes out on the RSG
material, causing it to fragment, and carrying the material far beyond
it would have otherwise traveled. This is important for the formation
of W-R bubbles - they may be composed in some cases mainly of RSG
material, perhaps material that has been dredged up. The W-R wind is
instrumental in dispersing this material over an area of tens of
parsecs, which the RSG wind by itself, with its low velocity, could
not accomplish.

The global structure of the bubble in multi-dimensions is not very
different from that predicted in one-dimension, but with considerable
fluctuations in the density and pressure of the interior. The various
instabilities and turbulence result in a Wolf-Rayet wind-termination
shock that is not spherical, but slightly elongated towards equatorial
latitudes. This has implications for the subsequent evolution of the
SN shock wave within this medium.

We note here that in some cases such as the one presented here, the
Wolf-Rayet wind-termination shock is not formed by the direct
interaction of the Wolf-Rayet wind with the surrounding medium, or
even with the wind from a pre-existing stage. Rather it results from
the Wolf-Rayet wind interaction with the MS shell, and a reflected
shock bouncing back until pressure equilibration with the ram pressure
of the freely expanding wind is achieved. This means that it is not
possible to predict the radius of this shock {\it a priori} from
wind-wind interaction, as is sometimes done nowadays for calculations
of the structure around gamma-ray bursts. This could lead to an
erroneous answer. {\em In calculating the radius of the Wolf-Rayet
wind-termination shock, one must take into account the previous
evolution of the circumstellar medium}.

Our results have important implications for the surroundings of
massive stars, and the environment in which supernovae, and possibly
gamma-ray bursts, evolve. We do point out that in our calculations we
have not considered the effects of the ionizing radiation from the
star. These were briefly detailed in Paper 1. It is possible that the
ionization front sweeping through the star may have a dynamical effect
that needs to be taken into account. We are now working on a code that
includes the effects of the ionization from the star. These
simulations will be detailed in forthcoming papers.

During the writing of this paper we have realized that the ionizing
effects in a similar case have recently been considered in simulations
by \citet{fhy06}. Unfortunately they do not provide details of the
shock structures in a one-dimensional model, which would have been
very useful to compare the direct effects of the ionization front. And
even though their highest resolution was larger than what we use here,
the fuzziness of their published figures precludes a detailed
comparison with this work. They see the formation of an ionization
front instability in the dense shell in the MS stage, but they do not
see in their simulations the Rayleigh-Taylor instabilities that are
mentioned herein in the RSG and W-R stages. Conversely they note the
formation of an R-T instability when the W-R wind passes over the
boundary between the RSG and MS wind. We do see the formation of
filamentary structure in this case (figure \ref{fig:bub2dzoom}, lower
left panel) but it is not clear that this is just due to the existing
instabilities that we have noted, the formation of a new R-T
instability, or even a combination of the two. Since they started with
a surrounding medium density different from that used here, the size
of the bubble and its properties would differ correspondingly, so it
is difficult to get a good read on how much the parameters such as
density, pressure and velocity are affected by the ionizing
radiation. The ionization front will raise the temperature, and
therefore the pressure, of the ionized region, which will affect the
evolution of the wind bubble in that region. One point of direct
comparison is the X-ray luminosity of the bubble over time, which can
be directly compared to Figure 19 of their paper. The variations in
the luminosity are much larger in our plot, possibly due to our more
approximate method. The luminosity calculated by us is slightly
larger, by a factor of a few, compared to their plot. This is partly
because we include all emission larger than 10$^6$K, whereas they
include only emission from 0.1-2.4KeV. However the overall similarity
in the plots, especially the large rise in luminosity during the W-R
phase, is striking. This also attests to the validity of our simpler
method of calculation.  In the final analysis, and without access to
their data, we conclude from their Figure 15 that while the formation
of an HII region plays a role, the pre-SN state of the gas at the end
of their calculation is not significantly different from that found
herein, and this is the main quantity that we are interested in.

Our most interesting results deal with the expansion of the SN shock
wave within this medium, which is the main goal of this paper. We find
several interesting effects. The evolution of the SN shock wave is
confined within this bubble for a substantially large period of time
of several doubling periods after the SN shock wave interacts with the
dense shell. This will often be the case for SNe that arise from W-R
stars (Type Ic SNe). The overall level of the emission from the
remnant is significantly reduced due to the lower density within the
cavity. Due to the fluctuations in density and pressure within the
bubble, there are several shocks and rarefaction waves seen
criss-crossing the remnant. A large range of velocities will be seen
at any given time throughout the remnant. The interior is almost
completely thermalized and heated to high, X-ray emitting temperatures
throughout. Thus, as suggested in Paper 1, SN shock waves in shells
could be one explanation for remnants which show centrally peaked
emission. However it must be pointed out that at least in this
particular case the density within the remnant is so low that the
emission measure in the interior is very small, and most X-ray
emission will be seen to arise only from the edge of the remnant.

The spherical SN shock wave interacting with an aspherical
wind-termination shock results in an aspherical, and considerably
wrinkled, transmitted shock. The corrugated nature of the shock wave
results in the interaction of the shock with the dense shell taking
place in a bit-by-bit fashion, with different parts of the shockwave
interacting with the dense shell at different times. As pointed out in
\S \ref{sec:sncsm1d}, the interaction of the shock wave with the shell
leads to a brightening up of the shell due to an increase in X-ray and
optical emission. In this case, as different parts of the shock wave
collide with the shell at different times, the shell will brighten up
in different places at different times, almost like blinking Christmas
lights.

This effect is reminiscent of the situation in SN 1987A. Optical
observations have shown the presence of very bright "hot spots" on the
equatorial ring surrounding the SN. The first spot was seen around
1997, and gradually over the next few years many more have been
visible. The latest HST pictures show spots almost all the way around
the ring. The ring itself is known to be the equatorial waist of a
circumstellar bipolar nebula surrounding the progenitor star, and the
spots can be interpreted as the interaction of a wrinkled shock wave
with the ring. This could be one example of the kind of situation
outlined in the previous paragraph. We caution that our simulations
are not mean to represent the situation in SN 1987A, which is
considerably more complex, and whose progenitor was probably a much
lower mass B3Ia star. Furthermore, in the case of SN 1987A we know
that the ionization from the star is important in creating an HII
region interior to the dense shell \citep{cd95}. And that perhaps
there are fingers (instabilities) pushing inwards from the ring with
which the shock is interacting. But nevertheless, the similarities are
intriguing. Even though this particular simulation may not be
representative of SN 1987A, it shows that it is possible for the SN
shell interaction within a bubble to occur in a discontinuous fashion
due to a wrinkled shock wave. SN 1987A may be the rare case where it
is possible to investigate these effects.

Our results provide useful pointers for investigating SNRs in
wind-blown cavities. A case in point is the Oxygen-rich remnant RCW
86. For several reason, including the low $n_e t$ values, the faint
emission and the X-ray profiles, it has been suggested \citep{vkb97}
that the remnant was formed in a wind-blown cavity. Our simulations
could be useful in testing some of these theories and predicting the
future evolution. Another remnant which may have been formed in a
wind-blown cavity is G292.0+1.8. This remnant shows the interesting
presence of instabilities, identified by some authors as
Rayleigh-Taylor instabilities that arise from the initial explosion
\citep{ghw05}. We note that in our simulations, such filamentary
structures are also formed after the shock-shell interaction, while
the reverse shock is headed back into the remnant and the transmitted
shock is trapped in the dense shell. This may provide an alternative
explanation for the presence of such structures. In future papers we
hope to study individual remnants in much further detail.

Our simulations illustrate that it is possible to start with a
spherical shock wave from a SN and still end up with a highly
aspherical remnant due to the complexity of the surrounding
medium. And as noted we have not even taken global asymmetries in the
surrounding medium into account, as in the case of SN 1987A. Although
we caution against reading too much into the specific details of a
single calculation, it serves to illustrate the basic point that SN
evolution in wind blown cavities differs considerably from that in a
wind or a constant density medium, and that the structure and
evolution of the remnant, its dynamics and kinematics, may all change
as a result. Furthermore, the X-ray emission from the remnant evolving
in the low-density cavity may be considerably reduced, with occasional
periods of increased luminosity. It is therefore necessary to
investigate SN shock waves while taking an accurate picture of the
ambient medium into which it expands. This however is only possible if
stellar evolution calculations including the parameters of the
mass-loss from massive stars for every timestep during its evolution
are readily available. Fortunately, the situation is consistently
improving, and more and more such calculations are now being
published.

In future papers we will investigate further the evolution of SN shock
waves in environments created by the pre-SN star. We will look at the
effects of more massive stars, and especially rotating stars, where
strong rotation leads to the formation of a wind that is faster and
denser at the poles of the star as compared to the equator
\citep{md01, do02}. These environments can have considerable effect on
the evolution of the shock wave.

\acknowledgements

VVDs research is supported by award \# AST-0319261 from the National
Science Foundation, and by NASA through grant \# HST-AR-10649 awarded
by the Space Science Telescope Institute. We would like to acknowledge
several very constructive discussions with Roger Chevalier, especially
in identifying the various instabilities that have been seen. Comments
and suggestions from John Blondin, Thierry Foglizzo, and Robin
Williams have proved extremely useful. We would like to thank the 
anonymous referee for some very useful comments and suggestions.

\clearpage

\begin{figure}
\includegraphics[angle=90,scale=0.75]{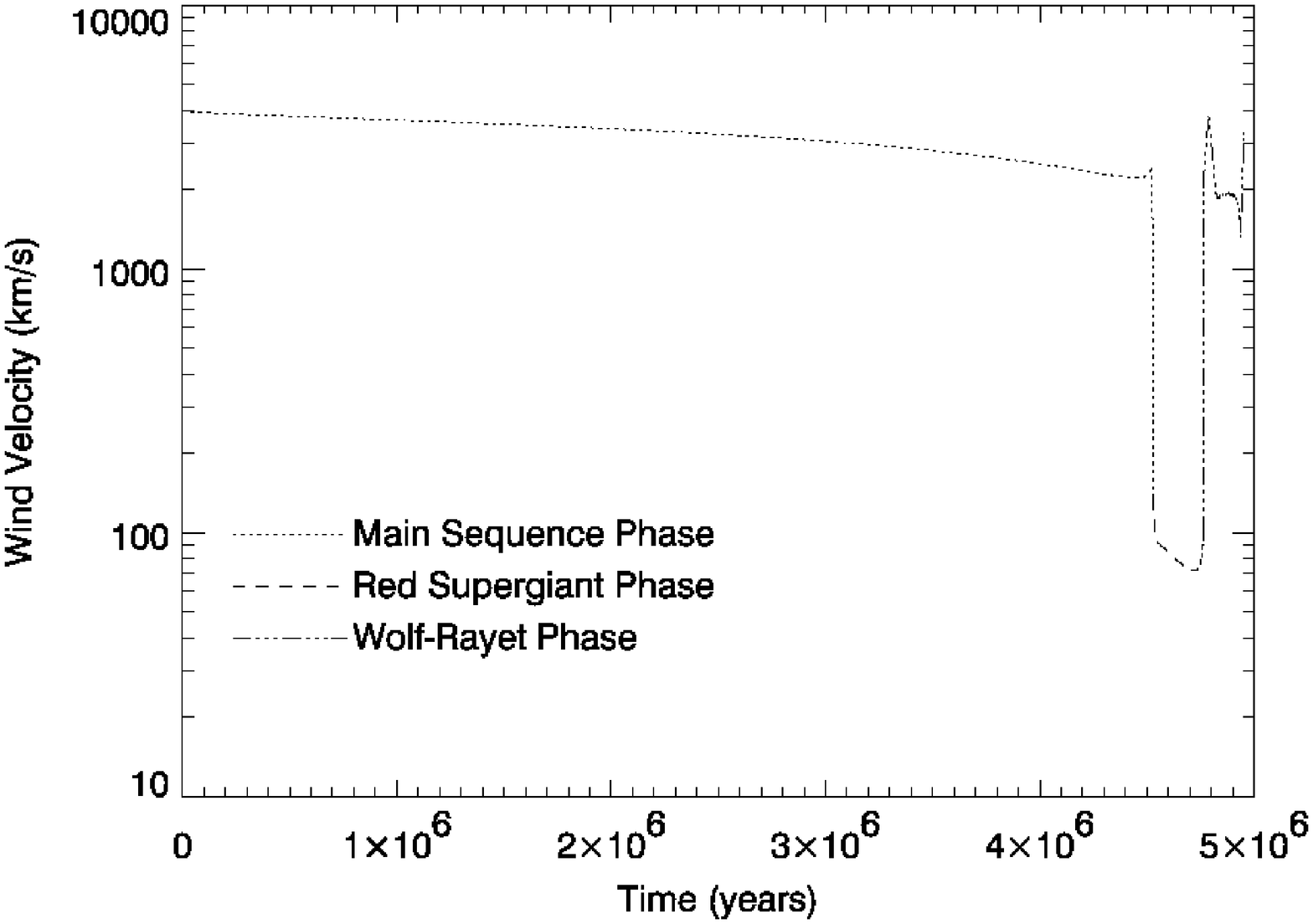}
\caption{Evolution of the velocity of the wind from the star with time}
\label{fig:bubvel}
\end{figure}

\begin{figure}
\includegraphics[angle=90,scale=0.75]{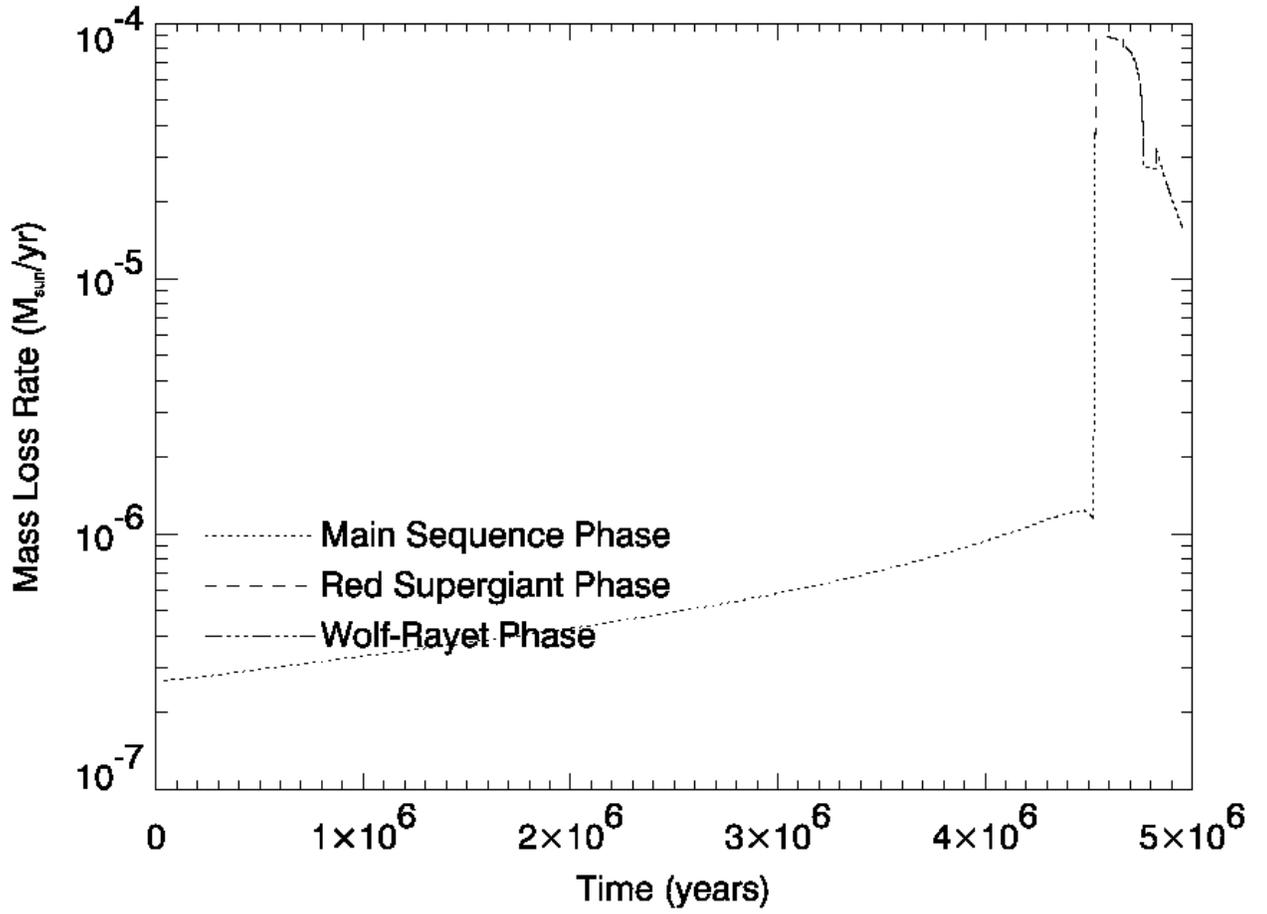}
\caption{Evolution of the mass-loss rate of the wind from the star with time}
\label{fig:bubmdot}
\end{figure}

\begin{figure}
\includegraphics{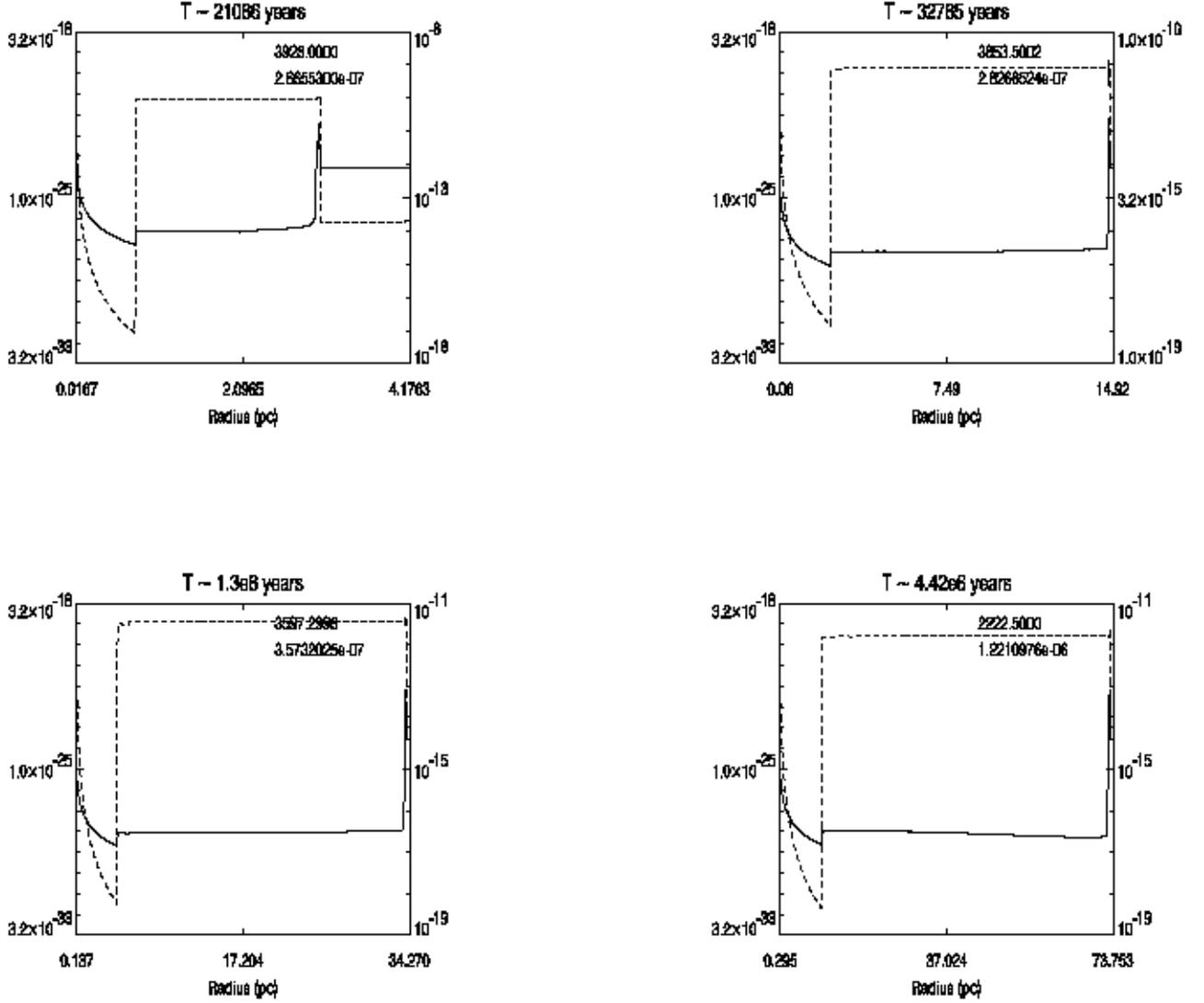}
\caption{Density and Pressure Profiles at various timesteps in the
evolution of the wind-blown nebula, during the Main Sequence
Stage. The solid line shows the density, the dashed lines the
pressure. Density scale is on LHS in g cm$^{-3}$, pressure on RHS in
cgs units. Time is given in years at the top of each panel. The two
numbers in the top right denote the velocity in km s$^{-1}$, and the
mass-loss rate in $\msun {\rm yr}^{-1}$. The X-axis scale is in
parsecs. Note that the grid is expanding with time. }
\label{fig:ms1d}
\end{figure}

\begin{figure}
\includegraphics{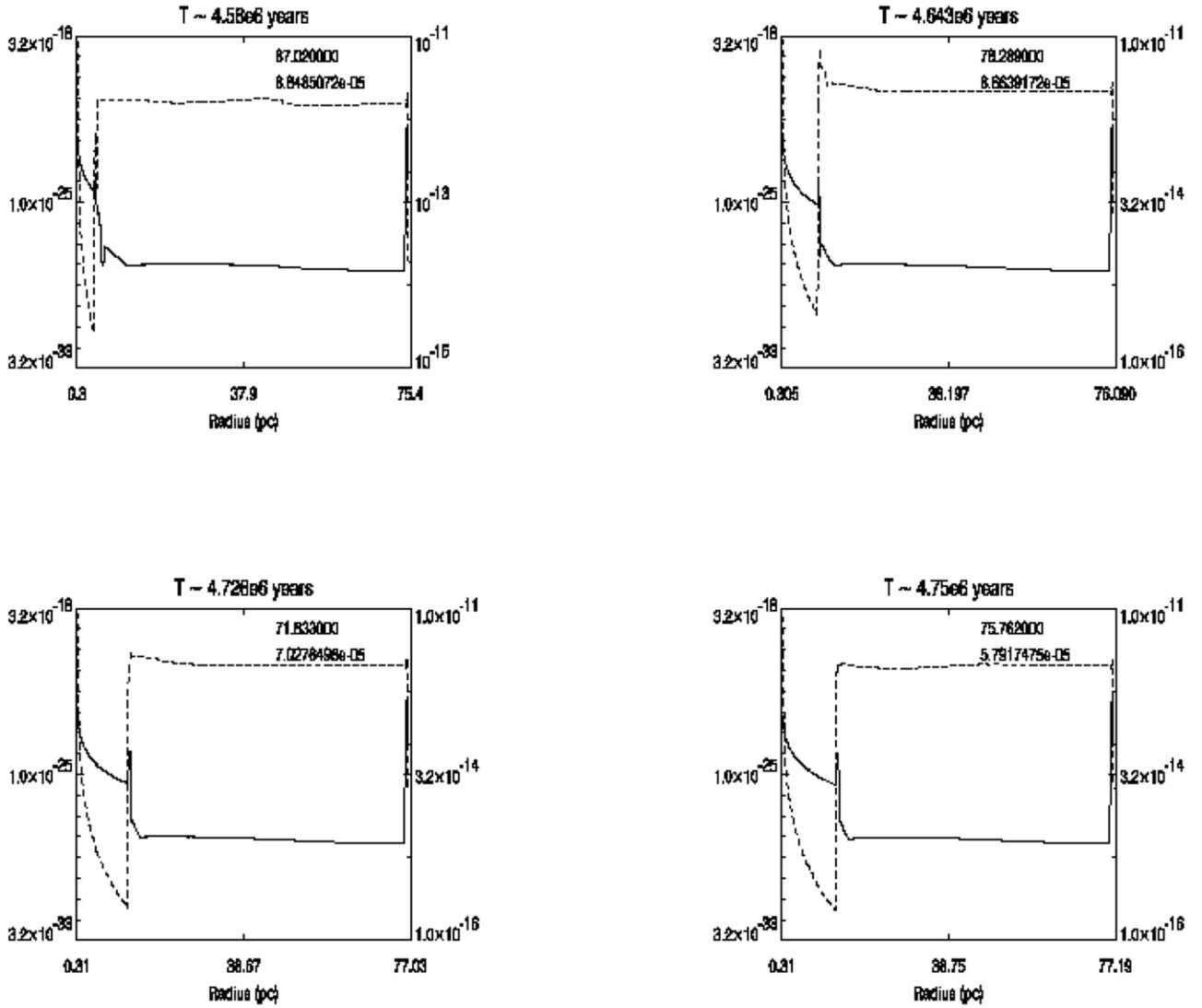}
\caption{Density and Pressure Profiles at various timesteps in the
evolution of the wind-blown nebula, during the Red Supergiant
Stage. Other details are as in Figure \ref{fig:ms1d}}
\label{fig:rsg1d}
\end{figure}

\begin{figure}
\includegraphics[scale=0.85]{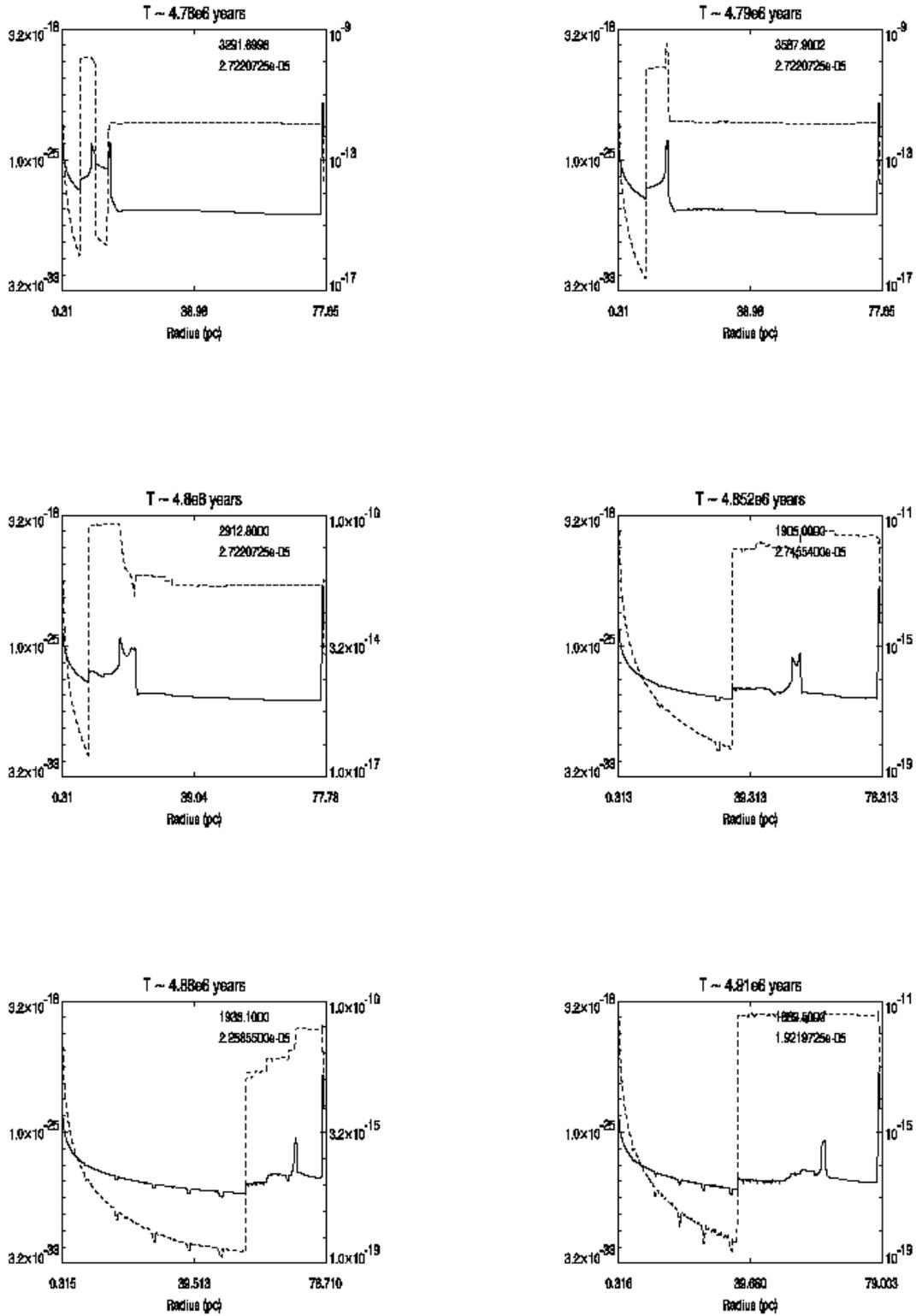}
\caption{Density and Pressure Profiles at various timesteps in the
evolution of the wind-blown nebula, during the Wolf-Rayet Stage. Other
details are as in Figure \ref{fig:ms1d}}
\label{fig:wr1d}
\end{figure}

\begin{figure}
\includegraphics[angle=90,scale=0.75]{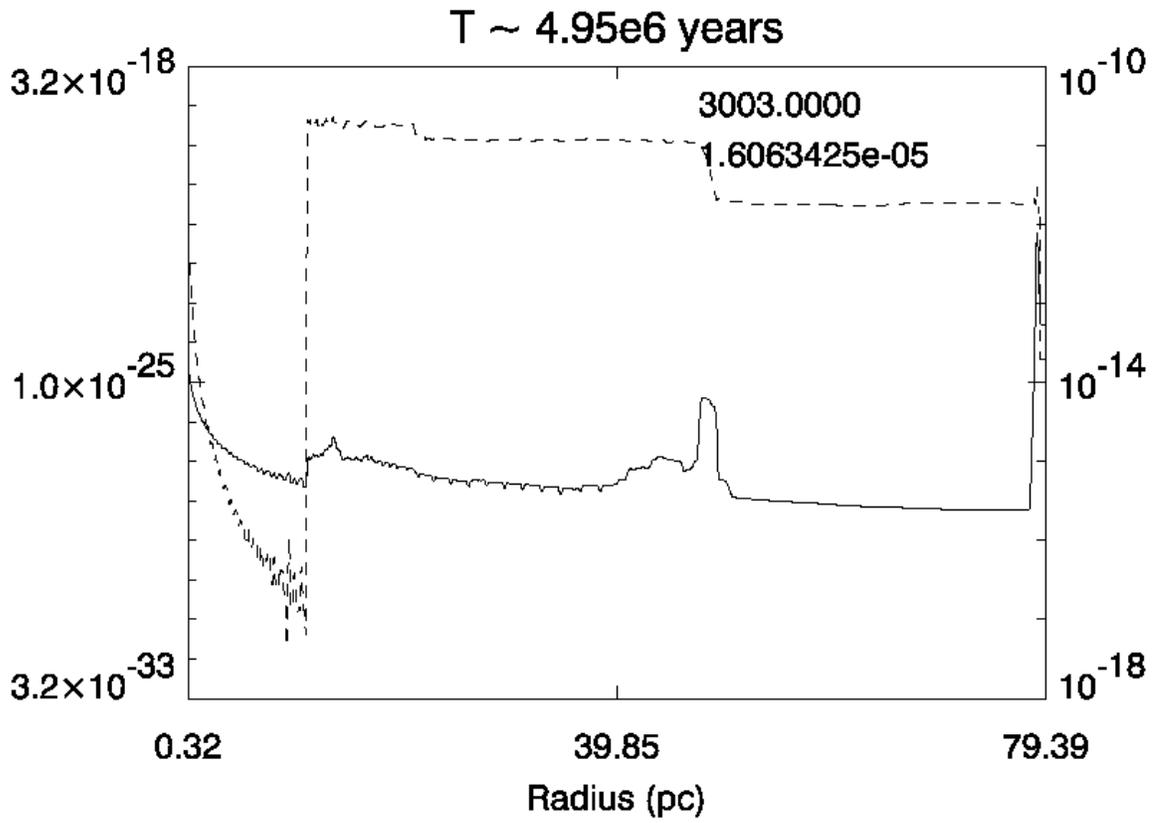}
\caption{Density and Pressure Profiles within the Circumstellar Bubble
at the end of the star's life, just prior to its death in a SN
explosion. Other details are as in Figure \ref{fig:ms1d}}
\label{fig:bub1dfinal}
\end{figure}

\begin{figure}
\includegraphics[angle=90,scale=0.75]{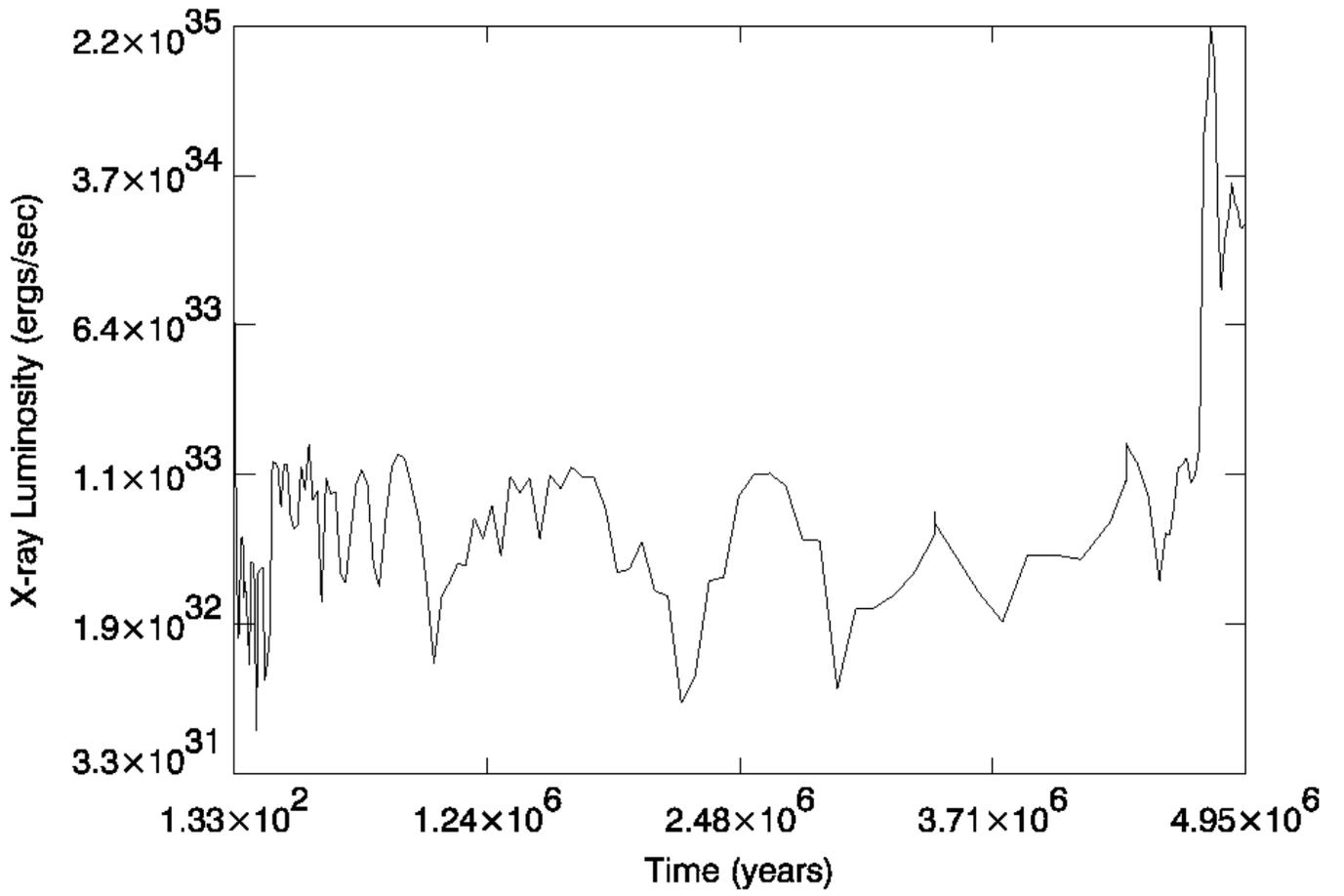}
\caption{The X-ray (thermal bremstrahhlung) luminosity during the evolution of the circumstellar wind-blown bubble.}
\label{fig:xlumbub}
\end{figure}

\begin{figure}
\includegraphics{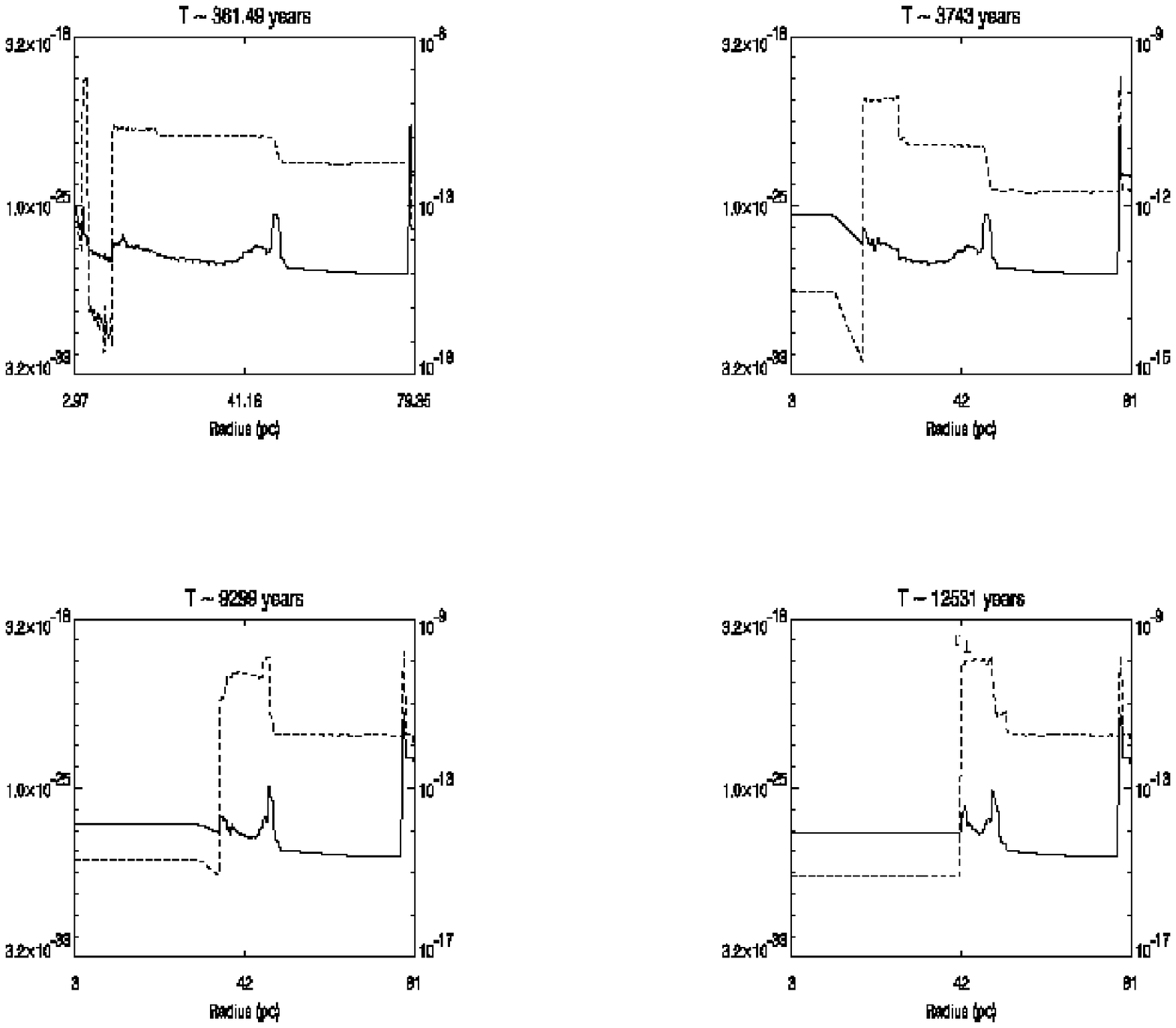}
\caption{Density and Pressure Profiles at various timesteps in the
evolution of the supernova remnant during the bubble. }
\label{fig:snbuba1d}
\end{figure}

\begin{figure}
\includegraphics{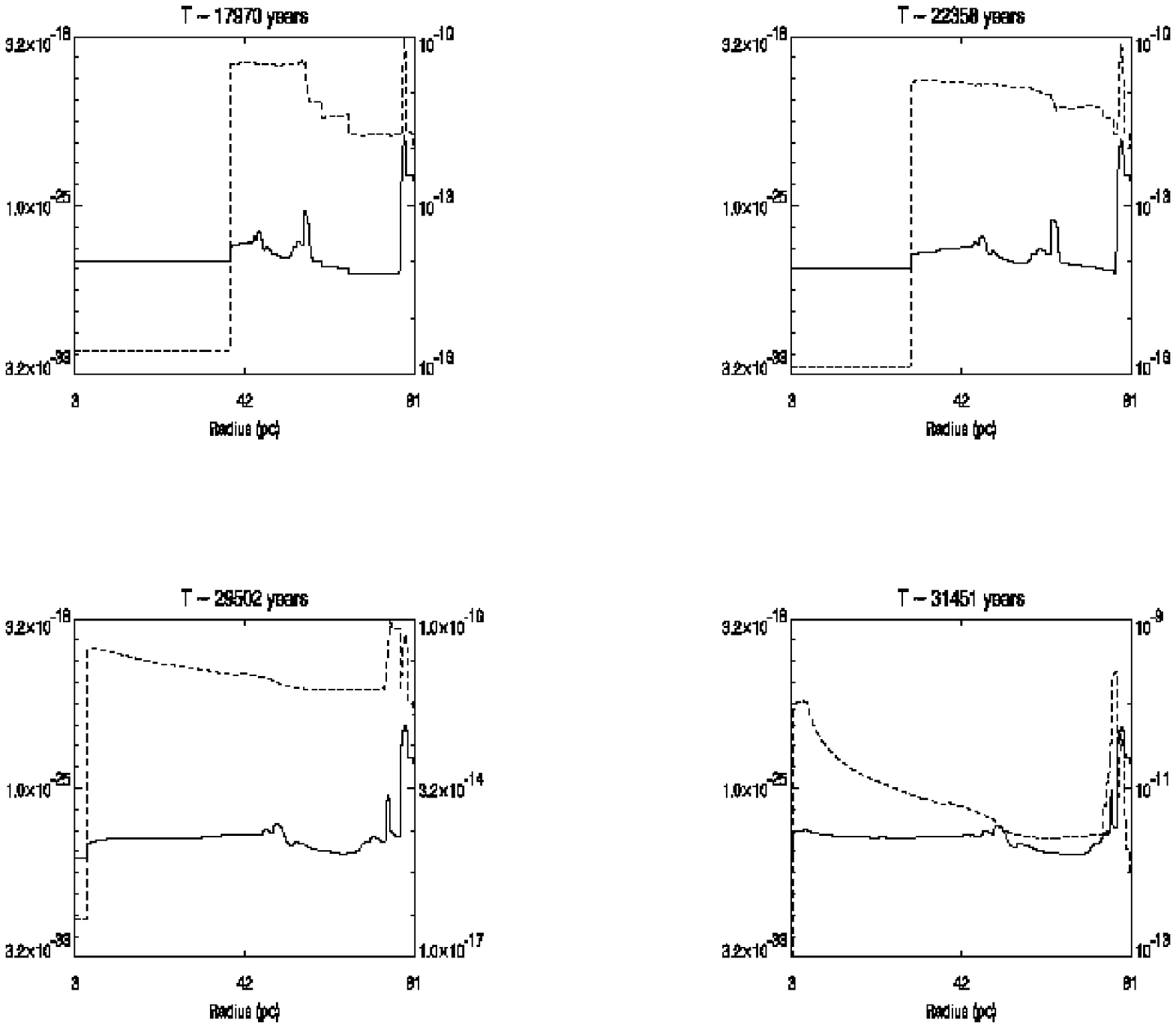}
\caption{Density and Pressure Profiles at various timesteps in the
evolution of the supernova remnant during the bubble.}
\label{fig:snbubb1d}
\end{figure}

\begin{figure}
\includegraphics{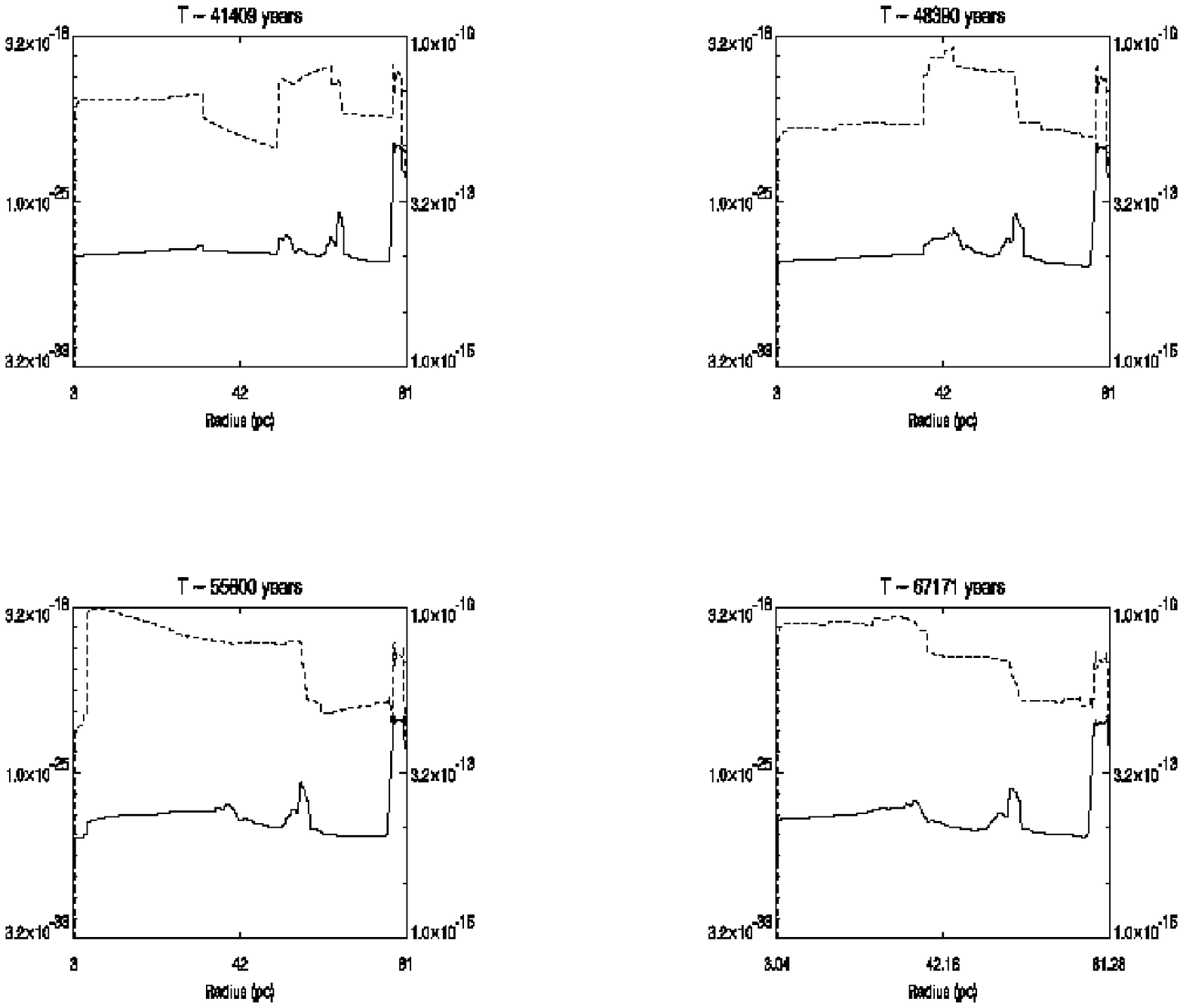}
\caption{Density and Pressure Profiles at various timesteps in the
evolution of the supernova remnant during the bubble.}
\label{fig:snbubc1d}
\end{figure}

\begin{figure}
\includegraphics{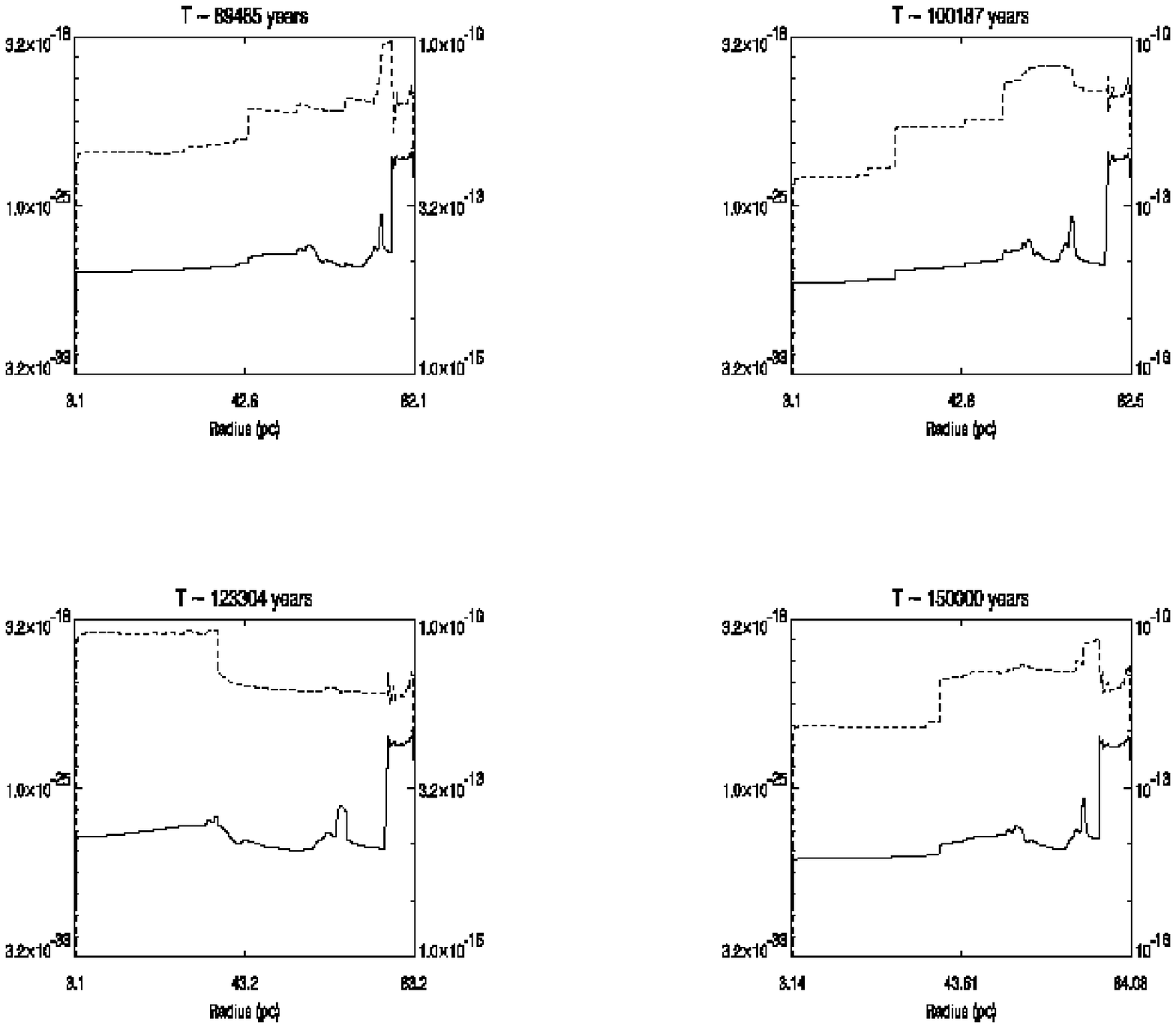}
\caption{Density and Pressure Profiles at various timesteps in the
evolution of the supernova remnant during the bubble.}
\label{fig:snbubd1d}
\end{figure}

\begin{figure}
\includegraphics[angle=90,scale=0.8]{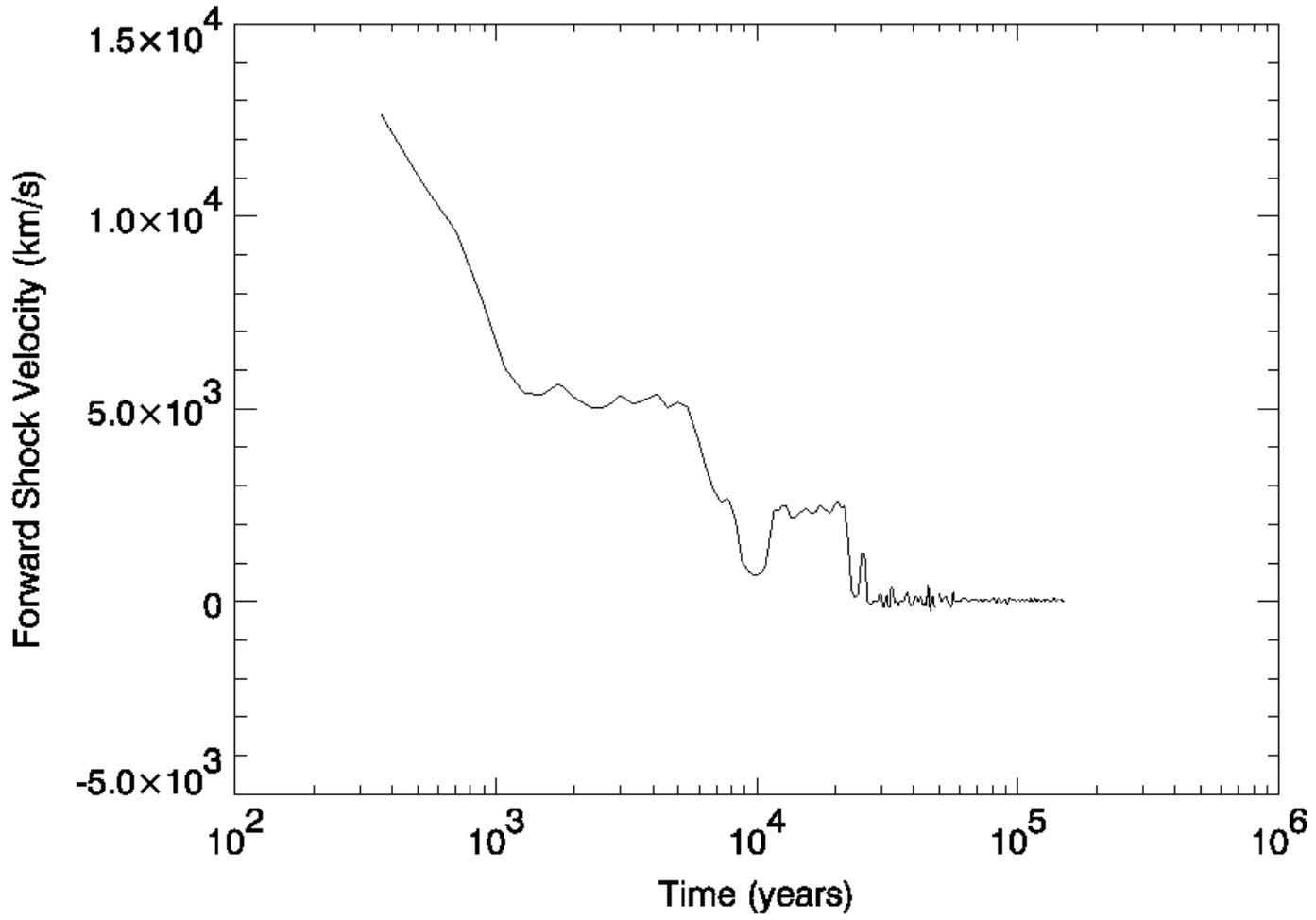}
\caption{The velocity of the forward shock with time over the
expansion of the remnant. }
\label{fig:snvel}
\end{figure}

\begin{figure}
\includegraphics[angle=90,scale=0.75]{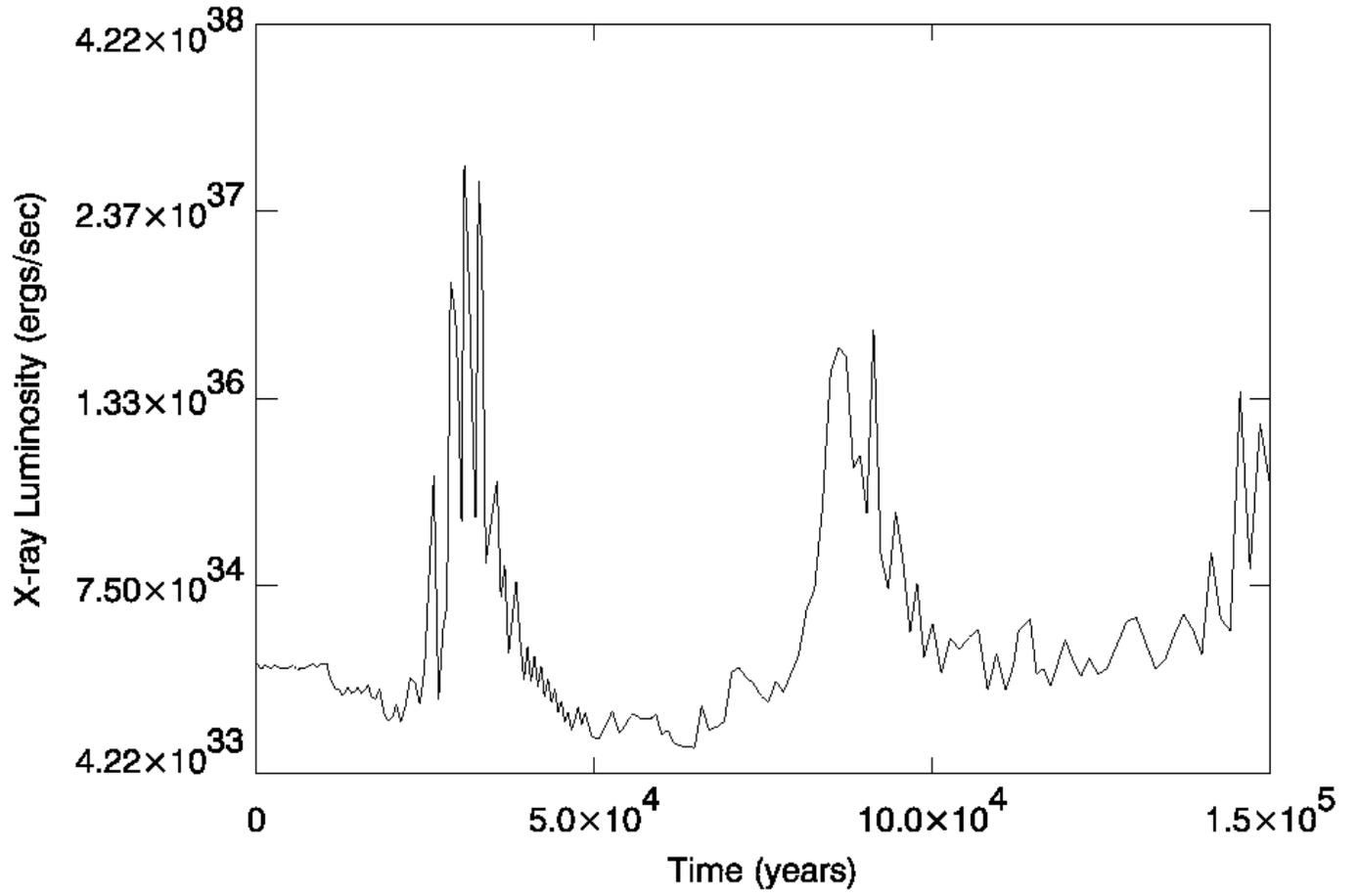}
\caption{The X-ray (thermal bremsstrahlung) luminosity during the SN evolution}
\label{fig:xlumsn}
\end{figure}

\begin{figure}
\includegraphics[scale=0.9]{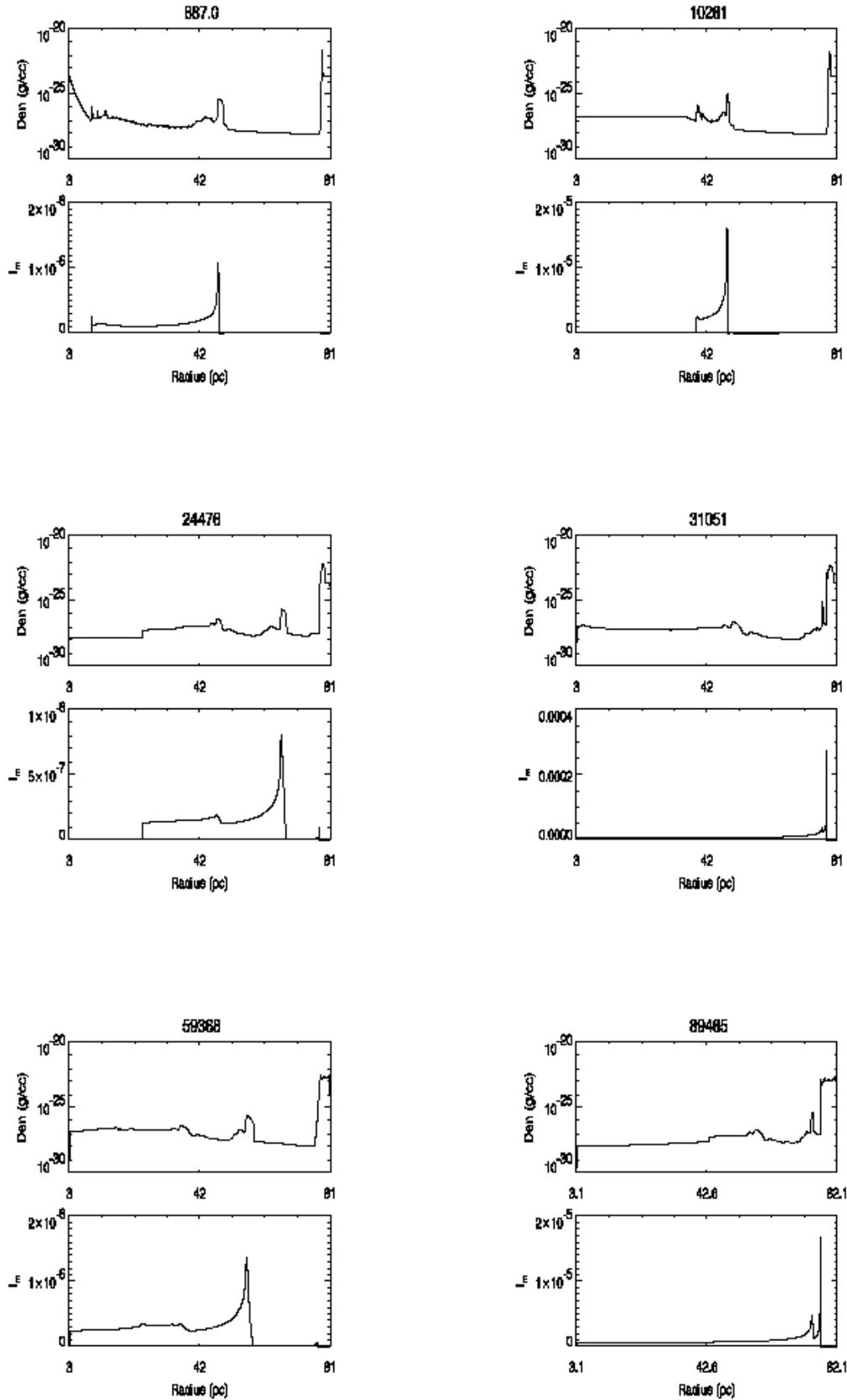}
\caption{The X-ray surface brightness profiles during the SN evolution}
\label{fig:xsbsn}
\end{figure}

\begin{figure}
\includegraphics[scale=0.9, clip=true]{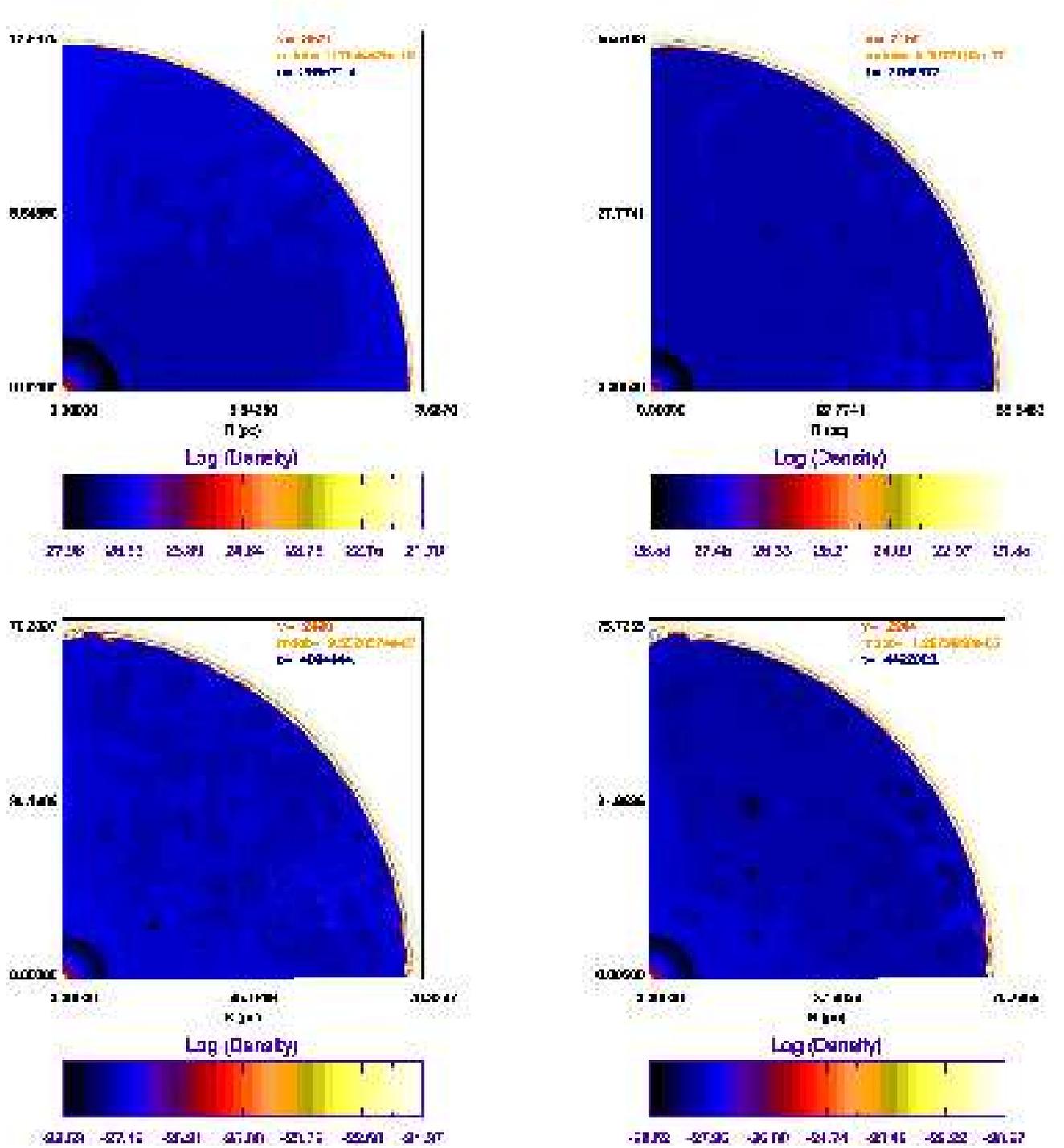}
\caption{Density plots of the two-dimensional evolution of the bubble
during the main-sequence stage. The numbers in the top right corner
give the wind parameters at each stage - velocity (in km s$^{-1}$),
mass loss rate (in $\msun {\rm}^{-1}$), and time (in years). }
\label{fig:bub2d_ms}
\end{figure}

\begin{figure}[t]
\includegraphics[angle=90,scale=0.85]{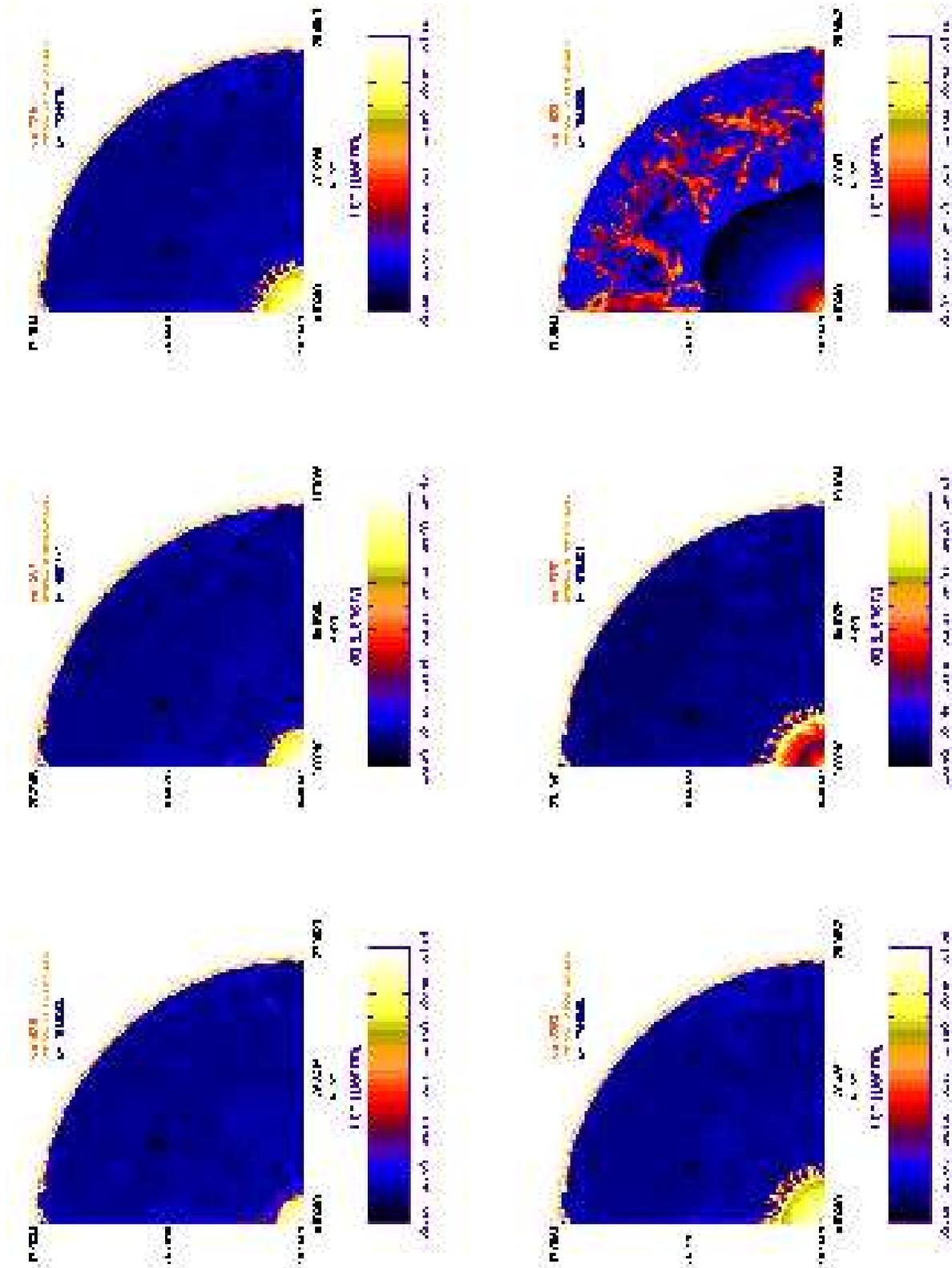}
\caption{The evolution of the wind-blown bubble in the red-supergiant
(first two panels) and Wolf-Rayet (panels three to six) phases. The
legend is as for Figure \ref{fig:bub2d_ms} }
\label{fig:bub2d_pms}
\end{figure}

\begin{figure}
\includegraphics[scale=0.95]{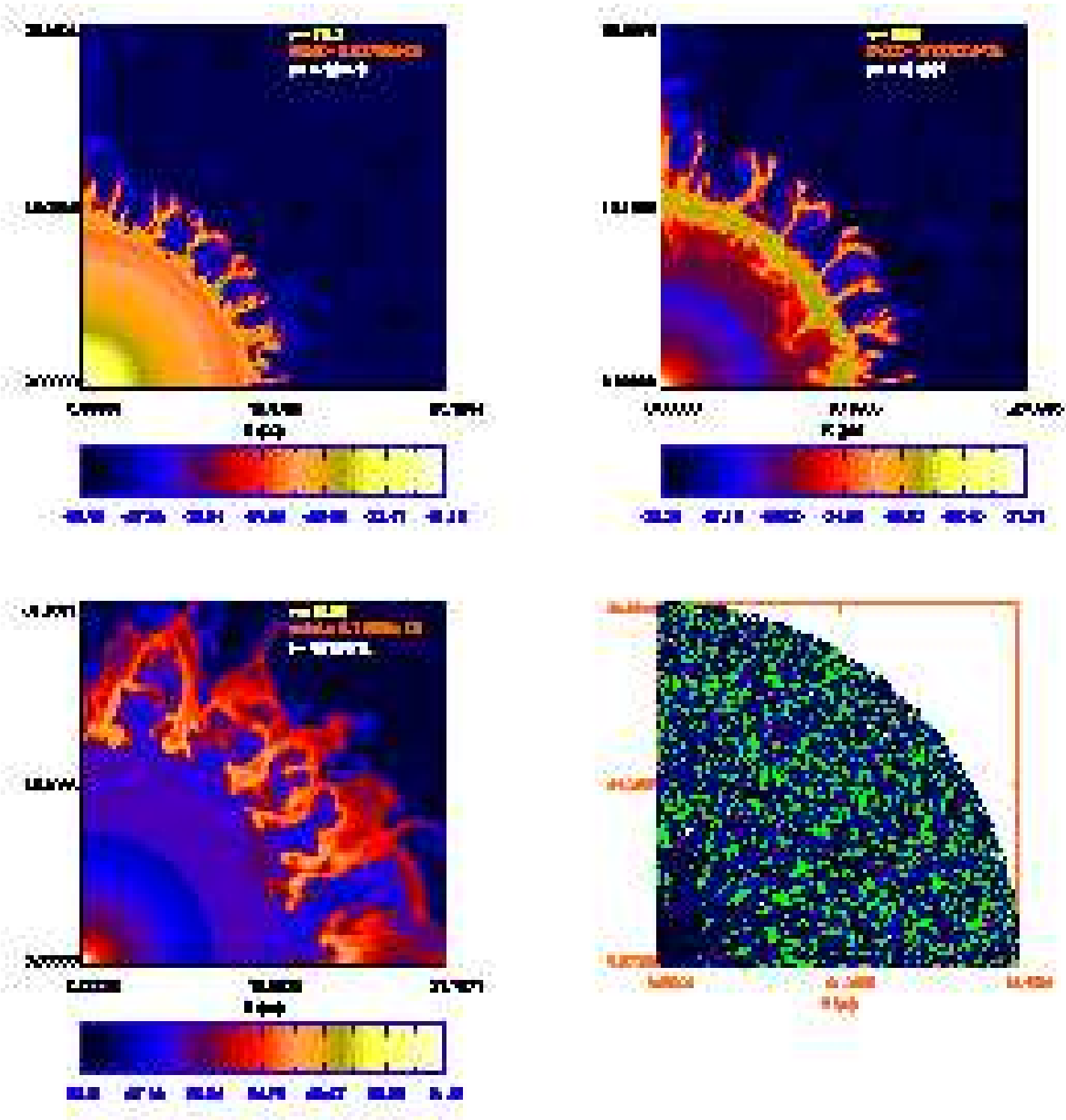}
\caption{4 frames emphasizing the bubble evolution at various phases,
especially to show the development of instabilities in those
phases. (a) Top left: The growth of Rayleigh-Taylor fingers in the Red
Supergiant wind (b) Top right: Growth of R-T instabilities in the
Wolf-Rayet wind (c) The interaction of the W-R and RSG winds, leading
to the fragmentation of the RSG wind (d) Velocity vectors plotted over
density contours at the end of the star's life. }
\label{fig:bub2dzoom}
\end{figure}

\clearpage

\begin{figure}[t]
\includegraphics*[angle=90, scale=0.85, origin=cr]{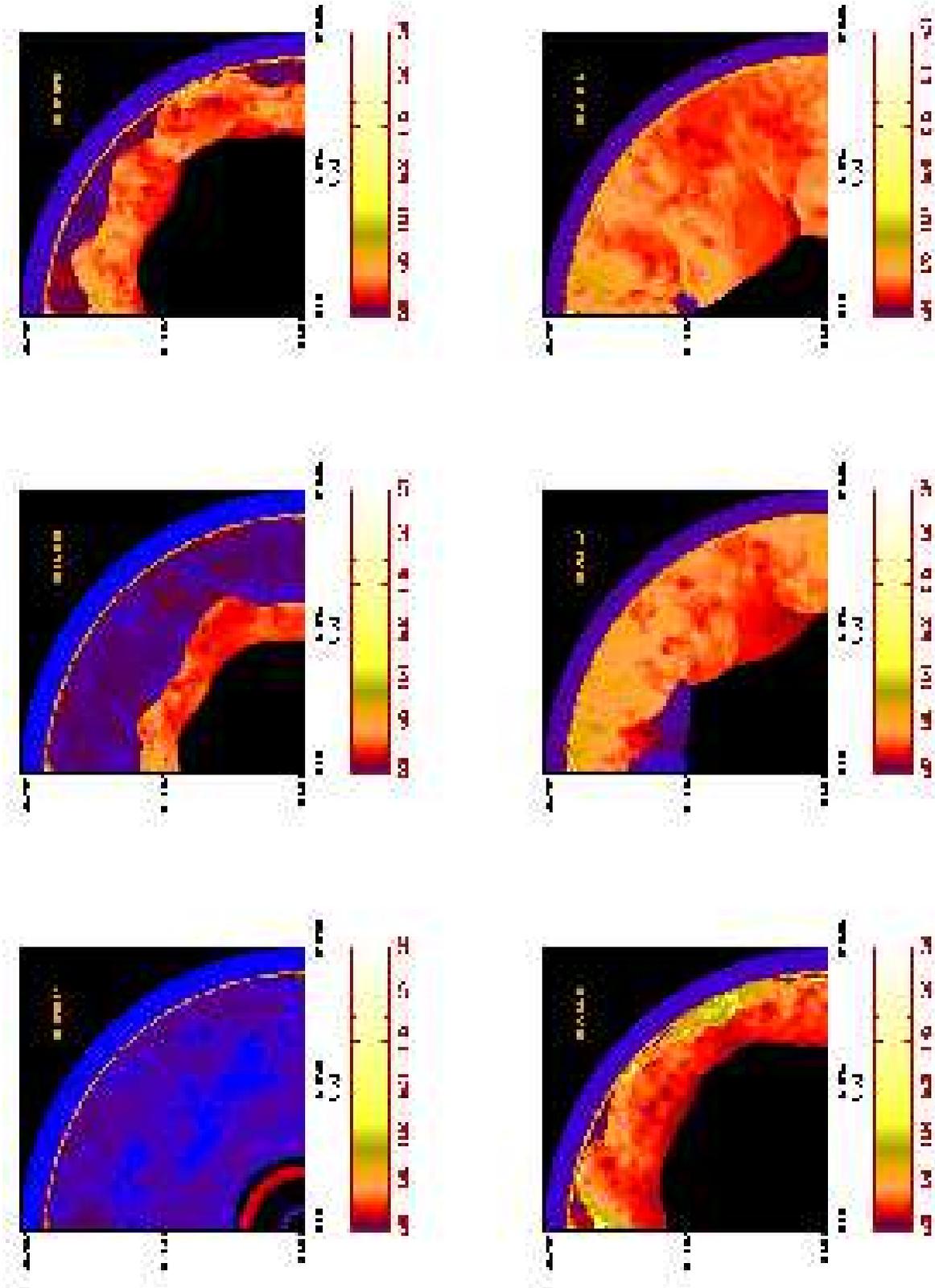}
\caption{These panels show the time-evolution of the supernova shock
wave within the wind-blown bubble. The panels show the pressure rather
than density as displayed in the previous figures. The time in years
is given at the top right of each plot. }
\label{fig:snbub2dpre}
\end{figure}

\clearpage

\begin{figure}[t]
\includegraphics*[angle=90, scale=0.85, origin=cr]{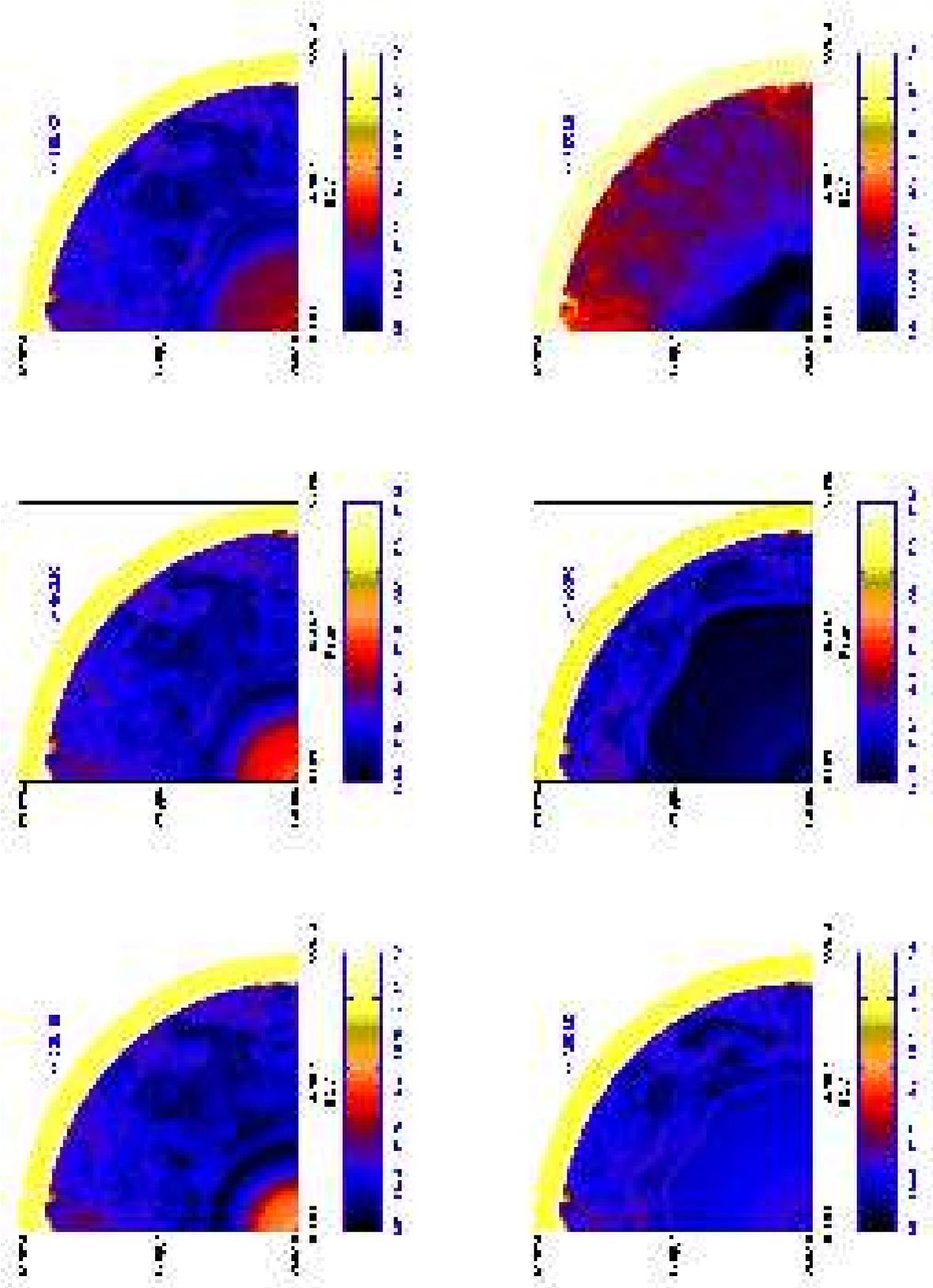}
\caption{Here we show the density evolution at the various timesteps
during the expansion of the SN shock wave. The panels clearly
demonstrate how the spherical shock wave becomes a wrinkled and
corrugated structure as it impacts the dense shell. }
\label{fig:snbub2dden}
\end{figure}


\clearpage

\begin{thebibliography}{}
\bibitem[Bandiera(1987)]{b87} Bandiera, R. 1987, \apj, 319, 885
Lundqvist, P. 1993, \apj, 405, 337
\bibitem[Blondin \& Lundqvist(1993)]{bl93} Blondin, J. M., \&
Lundqvist, P.~1993, \apj, 405, 337
\bibitem[Borkowski et al.(1996)]{bsbs96} Borkowski, K., Szymkowiak,
A. E., Blondin, J. M. \& Sarazin, C. L. 1996, \apj, 466, 866
\bibitem[Brighenti \& D'Ercole(1997)]{bd97} Brighenti, F., \&
D'Ercole, A.~1997, \mnras, 285, 387
\bibitem[Cappa et al.(2003)]{cec03} Cappa, C. E., Arnal, E. M.,
Cichowolski, S., Goss, W. M., Pineault, S, in ``A Massive Star
Odyssey: From Main Sequence to Supernova'', IAUS 212, Eds. K. A. van
der Hucht, A. Herrero, \& C. Esteban, SF: ASP, 596
\bibitem[Chevalier \& Dwarkadas(1995)]{cd95} Chevalier, R. A. \&
Dwarkadas, V. V. 1995, \apj, 452, L45
\bibitem[Chevalier \& Fransson(1994)]{cf94} Chevalier, R. A. \&
Fransson, C. 1994, \apj, 420, 268
\bibitem[Chevalier \& Liang(1989)]{cl89} Chevalier, R. A. \&
Liang, E. P. 1989, \apj, 344, 332
\bibitem[Chevalier(1982)]{c82} Chevalier, R. A. 1982, \apj, 258, 790
\bibitem[Chu, Gruendl, \& Guerrero(2006)]{cgg06} Chu, Y.-H., Gruendl,
R.~A., \& Guerrero, M.~A.~2006, in: Proceedings of the "The X-ray
Universe 2005", Ed. A. Wilson., ESA SP-604, Volume 1, (Noordwijk: ESA
Publications Division), 363
\bibitem[Chu et al.(2003)]{cgggw03} Chu, Y.-H., Guerrero, M.~A.,
Gruendl, R.~A., García-Segura, G., \& Wendker, H. L.~2003, \apj, 599,
1189
\bibitem[Ciotti \& D'Ercole(1989)]{cd89} Ciotti, L., \& D'Ercole,
A. 1989, \aap, 215, 347
\bibitem[Colella \& Woodward(1984)]{cw84} Colella, P., \& Woodward,
P. R. 1984, J. Comput. Phys., 59, 264
\bibitem[Dwarkadas(2007b)]{d07b} Dwarkadas, V. V, 2007, in preparation
\bibitem[Dwarkadas(2007a)]{d07a} Dwarkadas, V. V, 2007, ApSS, 307, 153
\bibitem[Dwarkadas(2005)]{d05} Dwarkadas, V. V, 2005, \apj, 630, 892
\bibitem[Dwarkadas(2004)]{d04} Dwarkadas, V. V, 2004, in ``Cosmic
explosions in three dimensions : Asymmetries in supernovae and
gamma-ray bursts'', Eds P. Hoflich, P. Kumar and J. C. Wheeler,
(Cambridge:CUP), 274
\bibitem[Dwarkadas \& Owocki(2002)]{do02} Dwarkadas, V. V., \& Owocki,
S. P. 2002, \apj, 581, 1337
\bibitem[Dwarkadas(2002)]{d02} Dwarkadas, V. V, 2002, in
``Interacting Winds from Massive Stars'', ASP Conference Proceedings
160, eds A.~F. J. Moffat and N.~St-Louis, (San Francisco: ASP), 141
\bibitem[Dwarkadas(2001)]{d01} Dwarkadas, V. V, 2001, JKAS, 34, 243
\bibitem[Dwarkadas \& Balick(1998)]{db98} Dwarkadas, V. V., \& Balick,
B. 1998, \apj, 497, 267
\bibitem[Dwarkadas et al.(1996)]{dcb96} Dwarkadas, V. V., Chevalier,
R. A., \& Blondin, J. 1996, ApJ, 457, 773
\bibitem[Frank \& Mellema(1994)]{fm94} Frank, A., \& Mellema, G. 1994, \aap, 
289, 937
\bibitem[Freyer, Hensler \& Yorke(2006)]{fhy06} Freyer, T., Hensler, G., \& Yorke, H.~W.~2006, \apj, 638, 262
\bibitem[Garcia-Segura \& Maclow(1995)]{gm95} Garcia-Segura, G.,
\& MacLow, M.-M. 1995, \apj, 455, 160
\bibitem[Garcia-Segura et al.(1996b)]{glm96b} Garcia-Segura, G.,
Langer, N., \& MacLow, M.-M. 1996, \aap, 316, 133
\bibitem[Garcia-Segura et al.(1996a)]{glm96a} Garcia-Segura, G.,
MacLow, M.-M., \& Langer, N.~ 1996, \aap, 305, 229
\bibitem[Ghavamian, Hughes, \& Williams(2005)]{ghw05} Ghavamian, P.,
Hughes, J.~P., \& Williams, T.~B.~2005, \apj, 635, 365
\bibitem[Hammer et al.(2006)]{hfs06} Hammer, F., Flores, H., Schaerer,
D., Dessauges-Zavadsky, M., Le Floc'h, E., Puech, M.~2006, A\&A, 454,
103
\bibitem[Hughes(1987)]{h87} Hughes, J. P. 1987, \apj, 314, 103
\bibitem[Koo \& Heiles(1995)]{kh95} Koo, B.-C., \& Heiles, C. 1995,
\apj, 442, 679
\bibitem[Koo \& McKee(1992a)]{km92a} Koo, B.-C., \& McKee, C. F. 1992a, \apj,
388, 93
\bibitem[Koo \& McKee(1992b)]{km92b} Koo, B.-C., \& McKee, C. F. 1992b,
\apj, 388, 103
\bibitem[Kudritzki et al.(1989)]{kppa89} Kudritzki, R. P., Pauldrach,
A., Puls, J., \& Abbott, D. C.~1989, \aap, 219, 205
\bibitem[Langer et al.(1994)]{lhl94} Langer, N., Hamann, W.-R.,
Lennon, M., Najarro, F., Pauldrach, A. W. A., Puls, J. 1994, \aap,
372, 9 290, 819
\bibitem[Langer(1989)]{l89} Langer, N. 1989, \aap, 220, 135
\bibitem[Langer(1994)]{l94} Langer, N. 1994, in ``Circumstellar Media
in the Late Stages of Stellar Evolution'', Proceedings of the 34th
Herstmonceaux conference, eds. R.E.S. Clegg, I. R. Stevens \&
W. P. S. Meikle, Cambridge: CUP, 1
\bibitem[Levenson et al.(1997)]{l97} Levenson, N. A. et al., 1997,
\apj, 484, 304
\bibitem[Luo \& McCray(1991)]{lm91} Luo, D., \& McCray, R.~1991, \apj, 379, 659
\bibitem[Maeder \& Desjacques(2001)]{md01} Maeder, A., \& Desjacques,
V.~2001, \aap, 372, 9
\bibitem[McKee(2004)]{m04}McKee, C. F. 2004, in ``Star Formation in
the Interstellar Medium: In Honor of David Hollenbach, Chris McKee and
Frank Shu'', ASP Conference Proceedings, Vol. 323, eds. D.~Johnstone,
F.C. Adams, D.N.C. Lin, D.A. Neufeld, and E.C. Ostriker, (San
Francisco: Astronomical Society of the Pacific), 21
\bibitem[Mdzinarishvili \& Chargeishvili(2005)]{mc05} Mdzinarishvili
T. G., \& Chargeishvili K. B. 2005, A\&A, 431, L1
\bibitem[Mellema(1995)]{m95} Mellema, G.~1995, \mnras, 277, 173
\bibitem[Mellema \& Frank(1995)]{mf95} Mellema, G., \& Frank, A. 1995, \mnras, 
273, 401
\bibitem[Nieuwenhuijzen \& de Jager(1990)]{nj90} Nieuwenhuijzen, H., \& de
Jager, C.,~1990, \aap, 231, 134
\bibitem[Ostriker \& McKee(1988)]{om88} Ostriker, J.~P., \& McKee,
C.~F.~1988, RvMP, 60, 10
\bibitem[Park et al.(2002)]{pr02} Park, S., Roming, P.~W.~A., et al.~2002, \apj, 564, L39
\bibitem[Raymond, Cox \& Smith(1976)]{rcs76} Raymond, J.~C., Cox,
D.~P., \& Smith, B.~W.~1976, \apj, 204, 290
\bibitem[Rozyczka et al.(1993)]{rtfb93} Rozyczka, M., Tenorio-Tagle,
G., Franco, J., \& Bodenheimer, P. 1993, \mnras, 261, 674
\bibitem[Ryan-Webber et al.(2004)]{rw04} Ryan-Weber, E. V., Meurer, G. R., Freeman, K. C., 
et al.~2004, \aj, 127, 1431
\bibitem[Sugerman et al.(2005b)]{sck05b}Sugerman, B.~E.~K., Crotts,
A.~P.~S., Kunkel, W.~E., Heathcote, S.~R., \& Lawrence, S.~S.~2005b,
\apjs, 159, 60
\bibitem[Sugerman et al.(2005a)]{sck05a}Sugerman, B.~E.~K., Crotts,
A.~P.~S., Kunkel, W.~E., Heathcote, S.~R., \& Lawrence, S.~S.~2005a,
\apj, 627, 888
\bibitem[Sutherland \& Dopita(1993)]{sd93} Sutherland, R., S., \&
Dopita, M. A. 1993, ApJS, 88, 253
\bibitem[Tenorio-Tagle et al.(1991)]{trfb91} Tenorio-Tagle, G.,
Rozyczka, M., Franco, J., \& Bodenheimer, P. 1991, \mnras, 251, 318
\bibitem[Tenorio-Tagle et al.(1990)]{tbf90} Tenorio-Tagle, G.,
Bodenheimer, P., Franco, J., \& Rozyczka, M. 1990, \mnras, 244, 563
\bibitem[Vink et al(1997)]{vkb97} Vink, J., Kaastra, J.~S., \&
Bleeker, J.~A.~M.~1997, \aap, 328, 628
\bibitem[Vishniac \& Ryu(1989)]{vr89} Vishniac, E. T., \& Ryu,
D.~1989, \apj, 337, 917
\bibitem[Vishniac(1983)]{v83} Vishniac, E. T.~1983, \apj, 274, 152
\bibitem[Weaver et al.(1977)]{wmc77} Weaver, R., McCray, R., Castor,
J., Shapiro, P., \& Moore, R. 1977, \apj, 218, 377
\bibitem[Wrigge et al.(2005)]{wcmw05} Wrigge, M., Chu, Y-H., Magnier,
E.~A., \& Wendker, H.~J.~2005,ApJ, 633, 248
\end{thebibliography}
\end{document}